\PassOptionsToPackage{normalem}{ulem} 

\documentclass[conference,final]{IEEEtran}

\usepackage{cite}
\usepackage[pdftex]{graphicx}
\graphicspath{{./plots/}{./diagrams/}}
\DeclareGraphicsExtensions{.pdf}

\usepackage[cmex10]{amsmath}
\usepackage{array}

\usepackage[caption=false,font=footnotesize]{subfig}
\usepackage{dblfloatfix}

\usepackage{hyperref}
\usepackage{xspace}
\usepackage[color=yellow,obeyFinal]{todonotes}
\usepackage{varioref}

\newcommand{\eg}{\textit{e.g.}\xspace}
\newcommand{\ie}{\textit{i.e.}\xspace}
\newcommand{\etal}{\textit{et al.}\xspace}

\usepackage{xcolor}
\definecolor{dark-red}{rgb}{0.4,0.15,0.15}
\definecolor{dark-blue}{rgb}{0.15,0.15,0.4}
\definecolor{medium-blue}{rgb}{0,0,0.5}
\hypersetup{
  colorlinks, linkcolor={dark-red},
  citecolor={dark-blue}, urlcolor={medium-blue}
}

\usepackage{balance}

\usepackage[boxed]{algorithm2e}

\SetStartEndCondition{ }{}{}%
\SetKwProg{Fn}{def}{\string:}{}
\SetKwFunction{Range}{range}
\SetKw{KwTo}{in}\SetKwFor{For}{for}{\string:}{}%
\SetKwIF{If}{ElseIf}{Else}{if}{:}{elif}{else:}{}%
\SetKwFor{While}{while}{:}{fintq}%
\AlgoDontDisplayBlockMarkers\SetAlgoNoEnd\SetAlgoNoLine%
\DontPrintSemicolon

\newcommand{\bl}{\mathcal{L}}
\newcommand{\bv}{\mathcal{O}}

\usepackage{changes}

\newcommand{\plotwidth}{.985\columnwidth}
\newcommand{\threeplotwidth}{.33\textwidth}

\begin{document}

\title{Revisiting Size-Based Scheduling\\with Estimated Job Sizes}

\author{
  \IEEEauthorblockN{Matteo Dell'Amico}
  \IEEEauthorblockA{EURECOM, France}
  \and
  \IEEEauthorblockN{Damiano Carra}
  \IEEEauthorblockA{University of Verona, Italy}
  \and
  \IEEEauthorblockN{Mario Pastorelli and Pietro Michiardi}
  \IEEEauthorblockA{EURECOM, France}
}

\maketitle

\begin{abstract}
We study size-based schedulers, and focus on the impact of inaccurate
job size information on response time and fairness. Our
intent is to revisit previous results, which allude to performance
degradation for even small errors on job size estimates, thus limiting
the applicability of size-based schedulers.

We show that scheduling performance is tightly connected to
workload characteristics: in the absence of large skew in the job size
distribution, even extremely imprecise estimates suffice to outperform
size-oblivious disciplines. Instead, when job sizes are heavily
skewed, known size-based disciplines suffer.

In this context, we show -- for the first time -- the dichotomy of
over-estimation versus under-estimation. The former is, in general, less
problematic than the latter, as its effects are localized to
individual jobs. Instead, under-estimation leads to severe problems
that may affect a large number of jobs.

We present an approach to mitigate these problems: our technique
requires no complex modifications to original
scheduling policies and performs very well. To support our claim, we
proceed with a simulation-based evaluation that covers an
unprecedented large parameter space, which takes into account a
variety of synthetic and real workloads.

As a consequence, we show that size-based scheduling is practical and
outperforms alternatives in a wide array of use-cases, even in
presence of inaccurate size information.

\end{abstract}

\section{Introduction}

In computer systems, several situations can be modeled as queues where
jobs (\eg, batch computations or serving data over the network) queue
to access a shared resource (\eg, processor or network). In this context,
size-based scheduling protocols, which prioritize jobs that are closest
to completion, are well known to have very desirable properties: the
shortest remaining processing time policy (SRPT) provides optimal mean
response time~\cite{schrage1966queue}, while the fair sojourn protocol
(FSP)~\cite{friedman2003fairness} provides similar efficiency while
guaranteeing strong fairness properties at the same time.

Despite these characteristics, however, scheduling policies similar to
SRPT or FSP are very rarely deployed in production: the \textit{de
facto} standard are instead policies similar to processor sharing
(PS), which divides resources evenly among jobs in the queue.  A key
reason is that, in real systems, job size is almost never
known \textit{a priori}. It is, instead, often possible to provide
\emph{estimations} of job size, which may vary in precision depending
on the use case; however, the impact of errors due to these
estimations in realistic scenarios is not yet well understood.

Perhaps surprisingly, very few works tackled the problem of size-based
scheduling with inaccurate job size information: as we discuss more in
depth in Section~\ref{sec:related}, the existing literature gives
somewhat pessimistic results, suggesting that size-based scheduling is
effective only when the error on size estimation is small; known
analytical results depend on restrictive assumptions on size
estimations, while simulation-based analyses only cover a limited
family of workloads. More importantly, no study we are aware of
tackled the design of size-based scheduling techniques that are
\emph{explicitly designed with the goal of coping with errors} in job size
information.

In Section~\ref{sec:errors}, we provide a qualitative analysis of the
impact of size estimation errors on the behavior of scheduling: we
show that, for heavy-tailed job size distributions, both FSP and SRPT
behave problematically when large jobs are under-estimated;
fortunately, it is possible to modify scheduling protocols to solve
this problem. The solution we propose is incarnated in FSPE+PS, a
simple modification to FSP. Analogous solutions can be applied to
protocols such as SRPT.

We developed a simulator, described in Section~\ref{sec:simulator}, to
study the behavior of FSP, SRPT, and FSPE+PS in a wide variety of
scenarios. Our simulator allows both replaying real traces and
generating synthetic ones varying system load, job size distribution
and inter-arrival time distribution; for both synthetic and real
workloads, scheduling protocols are evaluated on errors that range
between relatively small quantities and others that may vary even by
orders of magnitude.

From the experimental results of
Section~\ref{sec:experimental_results}, we highlight the
following ones, validated both on synthetic and real traces:
\begin{enumerate}
  \item When job size is not heavily skewed, SRPT and FSP outperform
    size-oblivious disciplines even when job size estimation is very
    imprecise and past work would hint towards important performance
    degradation; on the other hand, when the job size distribution is
    heavy-tailed, performance degrades noticeably;
  \item FSPE+PS does not suffer from the performance issues of FSP and
    SRPT; it provides good performance for a large part of the
    parameter space that we explore, being outperformed by a processor
    sharing strategy only when \emph{both} the job size distribution is heavily
    skewed \emph{and} size estimations are very inaccurate;
  \item FSPE+PS behaves fairly, guaranteeing that most jobs complete
    in an amount of time that is not disproportionate to their size.
\end{enumerate}

As we discuss in Section~\ref{sec:conclusion}, we conclude that our
work highlights and solves a key weakness of size-based scheduling
protocols when size estimation errors are present; the fact that
FSPE+PS consistently performs close to optimally highlights that
size-based schedulers are more viable in real systems than what was
known from the state of the art; we believe that our work can help
inspiring both the design of new size-based schedulers for real
systems and analytic research that can provide better insight on
scheduling when errors are present.

\section{Related Work}
\label{sec:related}

We discuss two main areas of related work: first, results for
size-based scheduling on single-server queues; second,
practical approaches devoted to the estimation of job sizes.

\subsection{Single-Server Queues}

Performance evaluation of scheduling policies in single-server queues
has been the subject of many studies in the last 40 years.  Among all
the policies, it has been shown that those that take into account the
size of the submitted job obtain the smallest mean response
time. Unfortunately, job sizes can often be only known approximatedly,
rather than exactly.  Since in our paper we consider this case, we
review the literature that targets this problem.

Perhaps surprisingly, not much work considers the effect of inexact
job size information on size-based scheduling policies. Lu
\etal~\cite{lu2004size} have been the first to consider this problem,
showing that size-based scheduling is useful only when job size
evaluations are reasonably good (high correlation, greater than 0.75,
between the real job size and its estimate). Their evaluation focuses
on a single heavy-tailed job size distribution, and does
  not explain the causes of the observed results. Instead, we show
the effect of different job size distributions (heavy-tailed,
memoryless and light-tailed), and we show how to modify the size-based
scheduling policies to make them robust to job estimation errors.

Wierman and Nuyens~\cite{wierman2008scheduling} provide analytical
results for a class of size-based policies, but consider an
impractical assumption: results depend on a bound on estimation error.
In the common case where most estimations are close to the real value
but there are outliers, bounds need to be set according to outliers,
leading to pessimistic predictions on performance. In our work,
instead, we do not impose any bound on the error.

To the best of our knowledge, these are the only works targeting job
size estimation errors for single-server queues. We remark that, by
using an experimental approach and replaying traces, we can take into
account phenomena that are not represented in the abstract M/G/1 or
G/G/1 models, such as periodic temporal patterns or correlations
between job size and submission time.

\subsection{Job Size Estimation}

In the context of distributed computational
  systems,
FLEX~\cite{wolf2010flex} and HFSP~\cite{pastorelli2013hfsp} proved
that size-based scheduling can perform well in practical
scenarios. In both cases, job size estimation is
  performed with very simple means (\ie, by sampling the execution
  time of a part of the job): such rough estimations are sufficient to
  provide good performance, and our results provide an explanation to this.

In several practical contexts, rough job size
  estimations are easy to perform. For instance, web servers can use
the size of files to serve as an estimator of job
size~\cite{schroeder2006web}, and the variability of the end-to-end
transmission bandwidth will determine the variability of the
estimation error. More elaborate job size estimation
  means are in several cases already available, since estimating job
  size is not only interesting for the scheduler; relevant examples
  are approaches that deal with predicting the size of MapReduce
  jobs~\cite{ARIA11, nsdi12-c, query_perf} and of database
  queries~\cite{lipton1995query}. The estimation error can be always
evaluated \textit{a posteriori}, and this evaluation can be used to
decide if the size-based scheduling works better than policies blind
to size.

\section{Scheduling Based on Estimated Sizes}
\label{sec:errors}

We now introduce formally the SRPT and FSP size-based scheduling
protocols, and describe the effects that estimation errors have on
their behavior, focusing on the difference between over- and
under-estimation. We notice that under-estimation triggers a behavior
which is problematic in particular for heavy-tailed job size
distributions, and we propose a solution to handle it.

\subsection{SRPT and FSP}
\label{sec:srpt_fsp}

The SRPT policy gives priority to the job with smallest remaining
processing time. SRPT is \emph{preemptive}: a new job with size
smaller than the remaining processing time of the running one will
preempt (\ie, interrupt) the currently running one. When the scheduler
has access to exact job sizes, SRPT has optimal mean sojourn time
(MST)~\cite{schrage1966queue} -- \emph{sojourn time}, or \emph{response
  time}, is the time that passes between a job's submission and its
completion.

SRPT may cause \emph{starvation} (\ie, never providing access to
resources): for example, if small jobs are constantly submitted, large
jobs may never get served. FSP (also known in literature as
\emph{fair queuing}~\cite{fair_queuing} and
\emph{Vifi}~\cite{gorinsky2007fair}) is a policy that doesn't suffer
from starvation by virtue of \emph{job aging}, \ie gradually
increasing the priority of jobs that are not scheduled. More
precisely, FSP serves the job that would complete earlier in a
\emph{virtual} emulated system running a processor sharing (PS)
discipline: since all jobs eventually complete in the virtual system,
they will also eventually be scheduled in the real one.

In the absence of errors, a policy such as FSP is particularly
desirable because it obtains a value of MST which is close to what is
provided by SRPT while guaranteeing a strong notion of fairness in the
sense that FSP \emph{dominates} PS: no jobs complete later in FSP than
in PS~\cite{friedman2003fairness}. When errors are present, such a
property cannot be guaranteed; however, as our experimental results in
Section~\ref{sec:experimental_results} show, FSP still preserves better
fairness than SRPT even when errors are present.

\subsection{Dealing With Errors: SRPTE and FSPE}
\label{sec:under_over}

We now consider the behavior of SRPT and FSP when the
  scheduler has access to \emph{estimated} job sizes rather than exact
  ones. For clarity, we will refer hereinafter to \emph{SRPTE} and
  \emph{FSPE} in this case.

\begin{figure}[!t]
  \centering
  \includegraphics[width=\columnwidth]{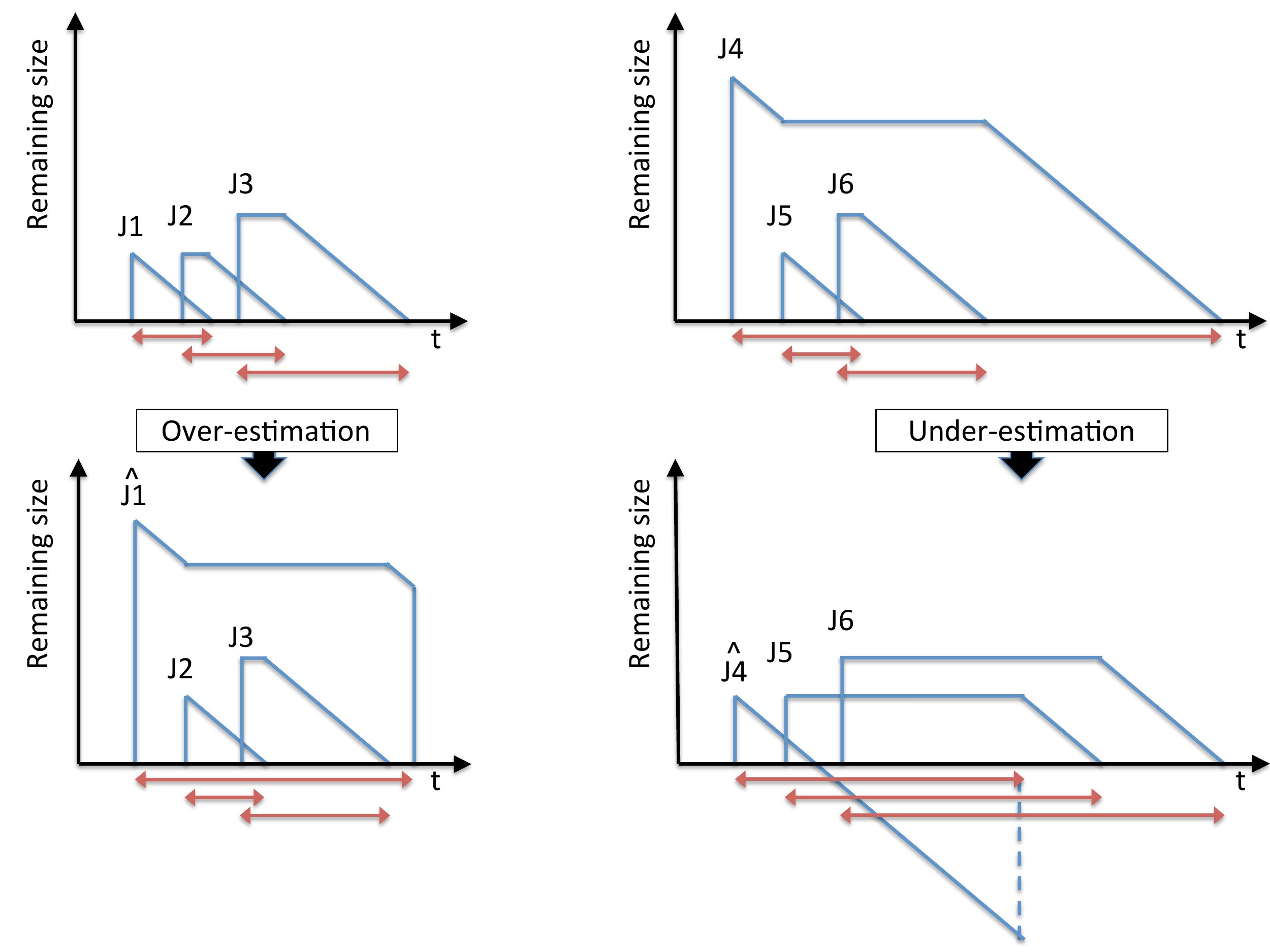}
  \caption{Examples of scheduling without (top) and with (bottom) errors.}
  \label{fig:under_over}
\end{figure}

In Figure~\vref{fig:under_over}, we provide an illustrative example
where a single job size is over- or under-estimated while the others
are estimated correctly, focusing (because of its simplicity) on
SRPTE; job sojourn times are represented by the horizontal arrows. The
left column of Figure~\ref{fig:under_over} illustrates the effect of
over-estimation. In the top, we show how the scheduler behaves without
errors, while in the bottom we show what happens when the size of job
$J1$ is over-estimated. The graphs shows the remaining (estimated)
processing time of the jobs over time (assuming a normalized service
rate of 1). Without errors, jobs $J2$ does not preempt $J1$, and $J3$
does not preempt $J2$.  Instead, when the size of $J1$ is
over-estimated, both $J2$ and $J3$ preempt $J1$. Therefore, the only
job suffering (\ie, experiencing higher sojourn time) is the one that
has been over-estimated. Jobs with smaller sizes are always able to
preempt an over-estimated job, therefore the basic property of SRPT
(favoring small jobs) is not significantly compromised.

The right column of Figure~\ref{fig:under_over} illustrates the effect
of under-estimation. With no estimation
errors (top), a large job, $J4$, is preempted by small ones ($J5$ and
$J6$). If the size of the large job is under-estimated (bottom), its
estimated remaining processing time eventually reaches zero: we call
\emph{late} a job with zero or negative estimated remaining processing
time. \emph{A late job cannot be preempted} by newly arrived jobs,
since their size estimation will always be larger than zero. In
practice, since preemption is inhibited, the under-estimated job
\emph{blocks the system} until the end of its service, with a negative
impact on multiple waiting jobs.

This phenomenon is particularly harmful when job sizes
  are heavily skewed: if the workload has few very large jobs and many
  small ones, a single late large job can significantly delay several
  small ones, which will need to wait for the late job to complete
  before having an opportunity of being served.

Even if the impact of under-estimation seems straightforward to
understand, surprisingly \emph{no work in the literature has ever
discussed it}. To the best of our knowledge, we are the
first to identify this problem, which significantly influences
scheduling policies dealing with inaccurate job size.

In FSPE, the phenomena we observe are analogous: job
  size over-estimation delays only the over-estimated job;
  under-estimation can result in jobs terminating in the virtual PS
  queue before than in the real system; this is impossible in absence
  of errors due to the dominance result introduced in
  Section~\ref{sec:srpt_fsp}. We therefore define \emph{late} jobs in
  FSPE as those whose execution is completed in the virtual system but
  not yet in the real one and we notice that, analogously to SRPTE, also in
  FSPE late jobs can never be preempted by new ones, and they block
  the system until they are all completed.

\subsection{Our Solution}
\label{sec:fspe+ps}

Now that we have identified the issue with existing size-based
scheduling policies, we propose our countermeasure. Several
alternatives are envisionable, including for example updating job size
estimations if new information becomes available as work progresses;
however, such a solution may not be always feasible, due to
limitations in terms of information or computational resources
available to the scheduler.

We propose, instead, a simple solution that requires no additional job
size estimation, based on the simple idea that \emph{late jobs should
  not prevent executing other ones}. This goal is achievable by
performing simple modifications to preemptive size-based scheduling
disciplines such as SRPT and FSP. The key property is that the
scheduler takes corrective actions when one or more jobs are
\emph{late}, guaranteeing that -- even when very large late jobs are
being executed -- newly arrived small jobs will get executed soon.

We show here \emph{FSPE+PS}, which is a modification to FSPE: the only
difference is that, when one or more jobs are late, (\ie, they have
completed in the emulated virtual system and not in the real one),
\emph{all late jobs are scheduled concurrently in a PS
  fashion}. FSPE+PS inherits from FSP and FSPE the guarantee that
starvation is absent, it is essentially as complex to implement as FSP
is and, as we show in Section~\ref{sec:experimental_results}, it
performs close to optimally in most experimental settings we observe.
Due to the dominance of FSP with respect to PS, if there are no size
estimation errors no jobs can ever become late: therefore, with no
error FSPE+PS is equivalent to FSP.

Several alternatives to FSPE+PS are possible: we experimented for
example with similar policies that are based on SRPT rather than on
FSP, that use a least-attained-service policy rather than a PS one for
late jobs, and/or that schedule aggressively jobs that are not late
yet as soon as at least one reaches the ``late'' stage. With respect
to the metrics we use in this work, their behavior is very similar to
the one of FSPE+PS, and for reasons of conciseness we do not report
about them here; we encourage the interested reader to examine their
implementation at \href{http://bit.ly/schedulers}{bit.ly/schedulers}.
In practice, most of the performance gain is due to the \emph{explicit
management of the late jobs}, and how late jobs are handled has no
significant impact on such a gain.

\begin{algorithm}[!t]
  \Fn{NextVirtualCompletionTime}{ 
    \lIf{$|\bv|=0$}{\Return $\emptyset$}
    \lElse{\Return $t + w_0 * |\bv|$}
  }
  \Fn{ProcessJob}{
    \lIf{$|\bl|\neq 0$}{\Return $\{(l_i, 1 / |\bl|) | l_i \in \bl \}$}
    \lElseIf{$|\bv|=0$}{\Return $\emptyset$}
    \Else{
      $k \leftarrow \min\{i | c_i\}$\;
      \Return $\{(j_k, 1)\}$
    }
  }
  \Fn{UpdateVirtualTime($s$)}{
    \lFor{$(\_, w_i, \_) \in \bv$}{
      $w_i \leftarrow w_i - (s - t) / |\bv|$\;
      $t \leftarrow s$
    }
  }
  \Fn{VirtualJobCompletion($s$)} {
    UpdateVirtualTime($\bv, t, s$)\;
    \lIf{$c_0$}{add $j_0$ to $\bl$}
    remove the first element from $\bv$\;
  }
  \Fn{RealJobCompletion($j$)}{
    find $i$ such that $j_i = j$\;
    $c_i \leftarrow$ False
  }
  \Fn{JobArrival($s, j, w$)}{
    UpdateVirtualTime($\bv, t, s$)\;
    insert $(j, w, \mathrm{True})$ in $\bv$ maintaining ordering
  }
  \caption{FSPE+PS.}
  \label{alg:fspe+ps}
\end{algorithm}

Algorithm~\vref{alg:fspe+ps} presents our implementation of FSPE+PS,
which is based on Friedman and Henderson's original description of
FSP~\cite[Section~4.4]{friedman2003fairness}. System state is kept in
three variables: the virtual PS queue state is kept in a list
$\bv$, containing $(j_i, w_i, c_i)$ tuples and ordered by the $w_i$
values: each such tuple represents a job $j_i$ having remaining
processing time $w_i$ in the virtual system, while the $c_i$ boolean
flag is set to True if $j_i$ is running in the real system; late jobs
are stored in a $\bl$ set; the variable $t$ stores the last time at
which the information in $\bv$ had been updated.

Computation is triggered by three events: if a job $j$ of estimated
size $w$ arrives at time $s$, JobArrival($s, j, w$) is called; when a
job $j$ completes, RealJobCompletion($j$) is called; finally, when a
job completes in the virtual system at time $s$,
UpdateVirtualTime($s$) is called (NextVirtualCompletionTime is used to
discover when to call VirtualJobCompletion). After each event, the
ProcessJob procedure is called to determine the new set of scheduled
jobs: its output is a set of $(j, s)$ pairs where $j$ is the job
identifier and $s$ is the fraction of system resources allocated to
it.

\section{Evaluation Methodology}
\label{sec:simulator}

Understanding size-based scheduling systems when there are estimation
errors is not a simple task. The complexity of the system makes an
analytical study feasible only if strong assumptions, such as a
bounded error~\cite{wierman2008scheduling}, are imposed. Moreover, to
the best of our knowledge, no analytical model for FSP (without
estimation error) is available, making an analytical evaluation of
FSPE and FSPE+PS even more difficult.

For these reasons, we evaluate our proposed scheduling policies
through simulation. The simulative approach is extremely flexible,
allowing to take into account several parameters -- distribution of the
arrival times, of the job sizes, of the errors. Previous simulative
studies (\eg,~\cite{lu2004size}) have focused on a subset of
these parameters, and in some cases they have used real traces. In our
work, we developed a tool that is able to both reproduce real traces
and generate synthetic ones. Moreover, thanks to the efficiency of the
implementation, we were able to run an extensive evaluation campaign,
exploring a large parameter space. For these
reasons, we are able to provide a broad view of the applicability of
size-based scheduling policies, and show the benefits and the
robustness of our solution with respect to the existing ones.

\subsection{Scheduling Policies Under Evaluation}
\label{sec:schedpolicies}

In this work, we take into account different scheduling policies, both size-based and blind to size. For the size-based disciplines, we consider SRPT as a reference for its optimality with respect to the MST. When introducing the errors, we evaluate SRPTE, FSPE 
and our proposal, FSPE+PS, described in Section~\ref{sec:errors}.

For the scheduling policies blind to size, we have implemented the
\emph{First In, First Out} (FIFO) and \emph{Processor Sharing} (PS)
disciplines. These policies are the default disciplines used in many
scheduling systems -- \eg, the default scheduler in
Hadoop~\cite{white2009hadoop} implements a FIFO policy, while Hadoop's
FAIR scheduler is inspired by PS; the Apache web server delegates
scheduling to the Linux kernel, which in turn implements a PS-like
strategy~\cite{schroeder2006web}. Since PS scheduling divides evenly
the resources among running jobs, it is generally considered as a
reference for its fairness (see the next section on the performance
metrics). Finally, we consider also the \emph{Least Attained Service}
(LAS)~\cite{rai2003analysis} policy. LAS scheduling, also known in the
literature as \emph{Foreground-Background}
(FB)~\cite{kleinrock1975theory} and \emph{Shortest Elapsed Time}
(SET)~\cite{coffman1973operating}, is a preemptive policy that gives
service to the job that has received the least service, sharing it
equally in a PS mode in case of ties. LAS scheduling has been designed
considering the case of heavy-tailed job size distributions, where a
large percentage of the total work performed in the system is due to
few very large jobs, since it gives more priority to small jobs than
what PS would do.

\subsection{Performance Metrics}
\label{sec:metrics}

We evaluate scheduling policies according to two main aspects:
\emph{mean sojourn time} (MST) and \emph{fairness}. 
Sojourn time is the time that passes between the
moment a job is submitted and when it completes; such a metric is
widely used in the scheduling literature.

The definition of fairness is more elusive: in his survey on the
topic, Wierman affirms that \textit{``fairness is an amorphous concept
that is nearly impossible to define in a universal
way''}~\cite{wierman2011fairness}. When the job size distribution is
skewed, it is intuitively unfair to expect similar sojourn times
between very small jobs and much larger ones; a common approach is to
consider \emph{slowdown}, \ie the ratio between a job's sojourn time
and its size, according to the intuition that the waiting time for a
job should be somewhat proportional to its size. In this work we focus
on the per-job slowdown, based on the intuition that as few jobs as
possible should experience ``unfair'' very high slowdown values;
moreover, in accordance with the definition by
Wierman~\cite{wierman2007fairness}, we also verify whether jobs having
a given size experience an ``unfair'' high expected slowdown value.

\subsection{Parameter Settings}
\label{sec:trace_generation}

\begin{table}[!t]
  \caption{Simulation parameters.}
  \label{table:parameters}
  \centering
  \begin{tabular}{|l|l|r|}
    \hline
    Parameter & Explanation & Default\\
    \hline
    \hline
    sigma & $\sigma$ in the log-normal error distribution & 0.5\\
    shape & shape for Weibull job size distribution & 0.25\\
    timeshape & shape for Weibull inter-arrival time & 1\\
    njobs & number of jobs in a workload & 10,000\\
    load & system load & 0.9\\
    \hline
  \end{tabular}
\end{table}

We empirically evaluate scheduling policies in a wide spectrum of
cases. Table~\vref{table:parameters} synthetizes the input parameters
of our simulator; they are discussed in the following.

\vspace{2mm}
\noindent
{\bf Job Size Distribution:} Job sizes are generated according to a
Weibull distribution, which allows us to evaluate both heavy-tailed
and light-tailed job size distributions. Indeed, the
\emph{\textbf{shape}} parameter allows to interpolate between
heavy-tailed distributions (shape $<1$), the exponential distribution
(shape$=1$), the Raleigh distribution ($\mathrm{shape}=2$) and
bell-shaped distributions centered around the `1' value
($\mathrm{shape}>2$). We set the \emph{scale} parameter of the
distribution to ensure that its mean is 1.

Since scheduling problems have been generally analyzed on heavy-tailed
workloads with job sizes using distributions such as Pareto, we
consider a default heavy-tailed case of $\mathrm{shape}=0.25$. In our
experiments, we vary the shape parameter between a very skewed
distribution with $\mathrm{shape}=0.125$ and a bell-shaped
distribution with $\mathrm{shape}=4$.

\vspace{2mm}
\noindent
{\bf Size Error Distribution:} We consider log-normally distributed
error values. A job having size $s$ will be estimated as $\hat{s}=sX$,
where $X$ is a random variable with distribution
\begin{equation}
\operatorname{Log-\mathcal{N}}(0,\sigma^{2}).
\label{eq:lognormal}
\end{equation}

This choice satisfies two properties: first, since error is
multiplicative, the absolute error $\hat s - s$ is proportional to the
job size $s$; second, under-estimation and over-estimation are equally
likely, and for any $\sigma$ and any factor $k > 1$ the (non-zero)
probability of under-estimating $\hat s\leq\frac s k$ is the same of
over-estimating $\hat s\geq ks$.  This choice also is substanciated by
empirical results: in our implementation of the HFSP scheduler for
Hadoop~\cite{pastorelli2013hfsp}, we found that the empirical error
distribution was indeed fitting a log-normal distribution.

The \emph{\textbf{sigma}} parameter
controls $\sigma$ in Equation~\ref{eq:lognormal}, with a default --
used if no other information is given -- of 0.5; with this value, the
median factor $k$ reflecting relative error is 1.40. In our
experiments, we let sigma vary between 0.125 (median $k$ is 1.088) and
4 (median $k$ is 14.85). 

It is possible to compute the correlation between the estimated and real
size as $\sigma$ varies. In particular, when sigma is equal to 0.5, 1.0, 2.0 and 4.0,  
the correlation coefficient is equal to 0.9, 0.6, 0.15 and 0.05 respectively.

The mean of this distribution is always larger than 1,
  and growing as sigma grows: the system is biased towards
  overestimating the aggregate size of several jobs, limiting the
  underestimation problems that FSPE+PS is designed
  to solve. Even in this setting, the results in
  Section~\ref{sec:experimental_results} show that the improvements
  obtained by using FSPE+PS are still significant.

\vspace{2mm}
\noindent
{\bf Job Arrival Time Distribution:} For the job inter-arrival time
distribution, we use a Weibull distribution for its flexibility to
model heavy-tailed, memoryless and light-tailed distributions.  We set
the default of its shape parameter (\emph{\textbf{timeshape}}) to 1,
corresponding to ``standard'' exponentially distributed arrivals. Also
here, timeshape varies between 0.125 (very bursty arrivals separated
by long intervals) and 4 (regular arrivals).

\vspace{2mm}
\noindent
{\bf Other Parameters:} The \emph{\textbf{load}} parameter is the mean arrival rate divided by
the mean service rate. As default
value, we use the same value of 0.9 used by Lu
\etal~\cite{lu2004size}; in our experiments we let the load parameter
vary between 0.5 and 0.999.

The number of jobs (\emph{\textbf{njobs}}) in each simulation round is
10,000 (in additional experiments -- not shown for space
  reasons -- we varied this parameter, without obtaining significant
  differences). For each experiment, we perform at least 30
repetitions, and we compute the confidence interval for a confidence
level of 95\%. For very heavy-tailed job size distributions (shape
$\leq 0.25$), results are very variable and therefore, in order to
obtain stable averages, we performed hundreds and/or thousands of
experiment runs, until the confidence levels have reached the 5\% of
the estimated values.

\subsection{Simulator Implementation Details}

Our simulator is available under the Apache V2
license at \url{https://bitbucket.org/bigfootproject/schedsim}.
It has been conceived with ease of prototyping in mind: for example,
our implementation of FSPE as described in Section~\ref{sec:errors}
requires 53 lines of code. Workloads can be both replayed from real
traces and generated synthetically.

The simulator has been written with a focus on computational
efficiency. It is implemented using an event-based paradigm, and we
used efficient data structures based on B-trees
(\href{http://stutzbachenterprises.com/blist/}{stutzbachenterprises.com/blist/}).
As a result of these choices, a ``default'' workload of 10,000 jobs is
simulated in around half a second, while using a single core in our
machine with an Intel T7700 CPU. We use IEEE 754 double-precision
floating point values to represent time and job sizes.

\section{Experimental Results}
\label{sec:experimental_results}

We now present our experimental findings. For all the results shown in
the following, the parameters whose values are not explicitly stated
take the default values shown in Table~\ref{table:parameters}. For
the readability of the figures, we do not show the confidence intervals: 
for all the points, in fact, we have
performed a number of runs sufficiently high to obtain a confidence
interval smaller than 5\% of the estimated value.
We first present our results on synthetic workloads generated
according to the methodology of Section~\ref{sec:trace_generation}; we
then show the results by replaying two real-world traces from
workloads of Hadoop and of a Web cache.

\subsection{Synthetic Workloads}
\label{sec:synthetic}
\label{sec:exp_slowdown}

\begin{figure*}[!t]
  \centering
  \subfloat[SRPTE.]{
    \includegraphics[width=\threeplotwidth]{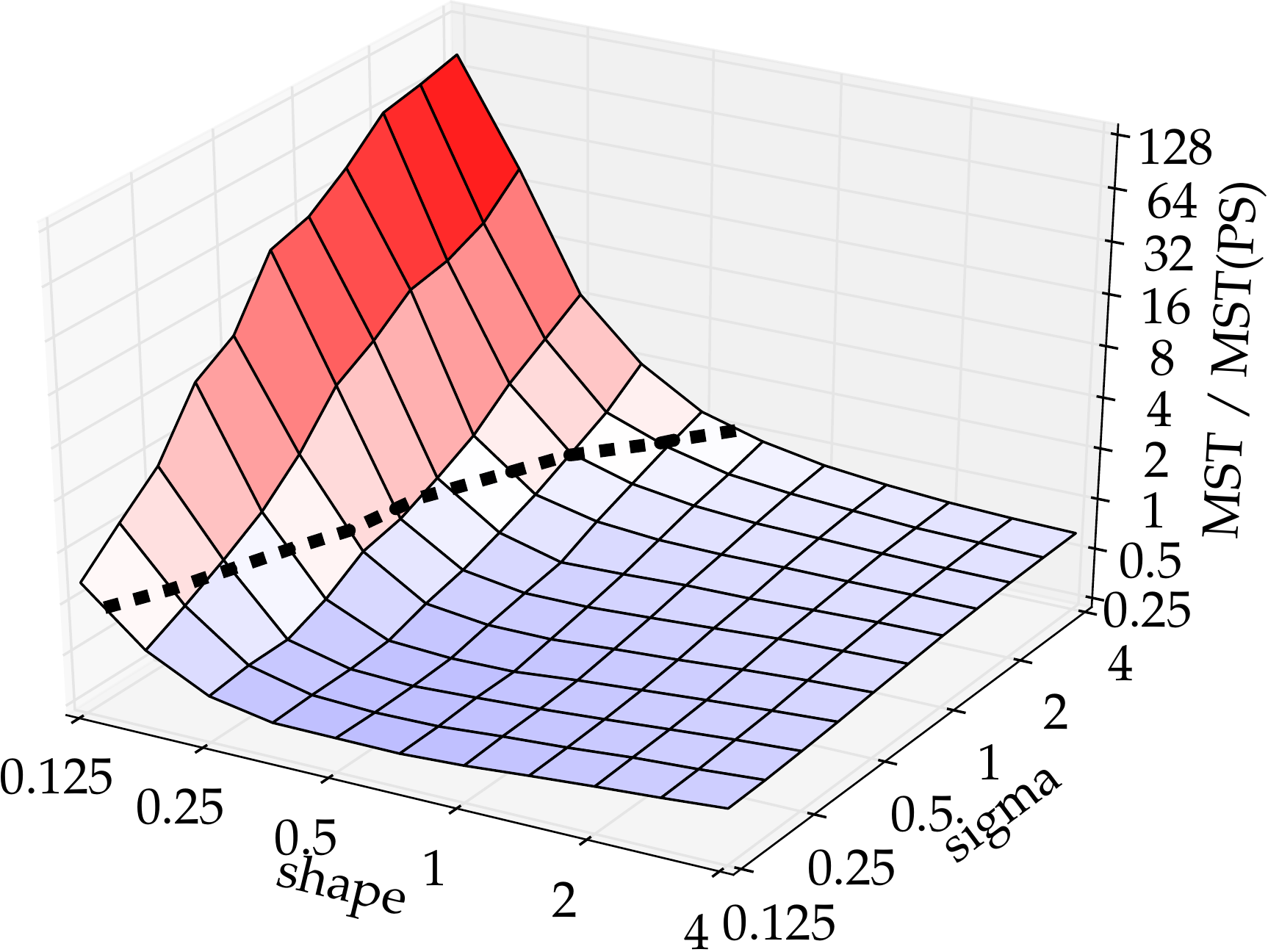}
    \label{fig:3d_srpte}
  }
  \subfloat[FSPE.]{
    \includegraphics[width=\threeplotwidth]{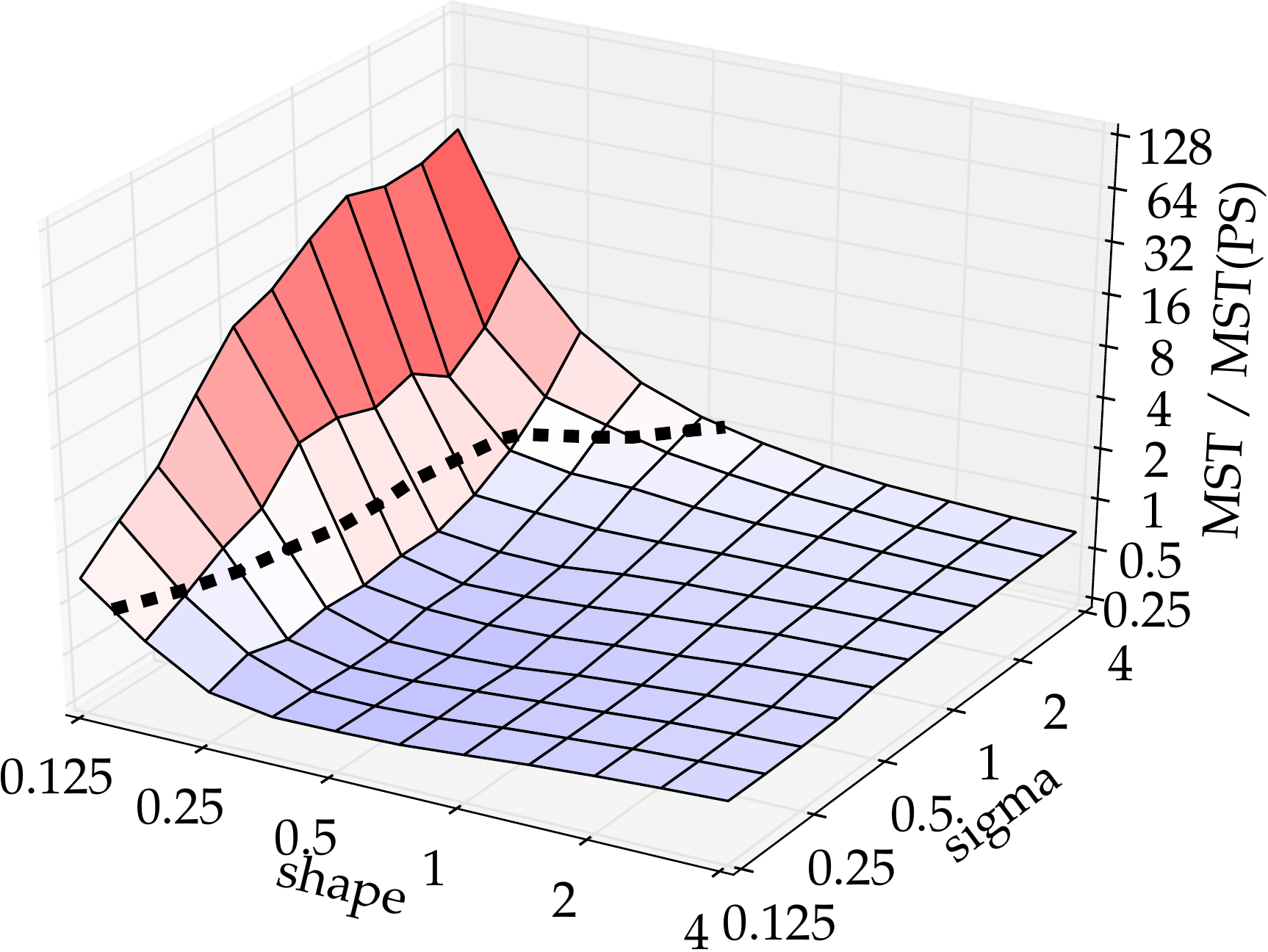}
    \label{fig:3d_fspe}
  }
  \subfloat[FSPE+PS.]{
    \includegraphics[width=\threeplotwidth]{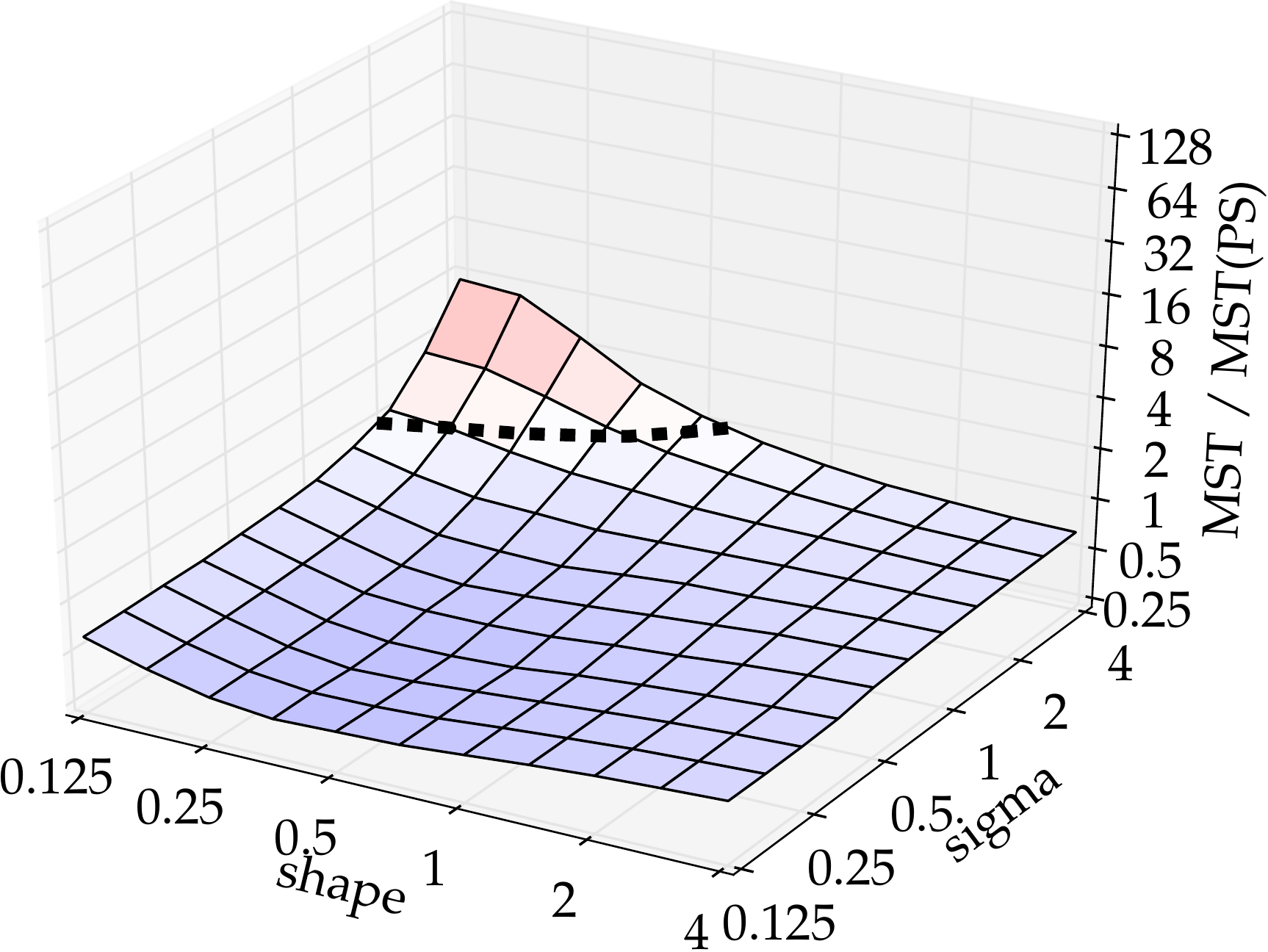}
    \label{fig:3d_fspe+ps}
  }
  \caption{Mean sojourn time against PS.}
  \label{fig:3d}
\end{figure*}

\noindent
{\bf Mean Sojourn Time Against PS:} We begin our analysis by comparing
the three size-based scheduling policies, using PS as a baseline
because PS and its variants are the most widely used set of scheduling
policies in real systems. In Figure~\ref{fig:3d} we plot the value of
the MST obtained using respectively SRPTE, FSPE and FSPE+PS,
normalizing it against the MST of PS. We vary the sigma and shape
parameters influencing respectively job size distribution and error
rate; we will see that these two parameters are the ones that
influence performance the most. Values lower than one (below the
dashed line in the plot) represent regions where size-based schedulers
perform better than PS.

In accordance with intuition and to what is known from the literature,
we observe that the performance of size-based scheduling policies
depends on the accuracy of job size estimation: as sigma grows,
performance suffers. In addition, from Figures~\ref{fig:3d_srpte}
and~\ref{fig:3d_fspe}, we observe a new phenomenon: \emph{job
  size distribution impacts performance even more than size estimation
  error.} On the one hand, we notice that large areas of the plots
($\textrm{shape} > 0.5$) are almost insensitive to estimation errors;
on the other hand, we see that MST becomes very large as job size skew
grows ($\textrm{shape} < 0.25$). We attribute this latter phenomenon
to the fact that, as we highlight in Section~\ref{sec:errors}, late
jobs whose estimated remaining (virtual) size reaches zero are never
preempted. If a large job is under-estimated and becomes \emph{late} with
respect to its estimation, small jobs will have to wait for it to
finish in order to be served.

As we see with Figure~\ref{fig:3d_fspe+ps}, \emph{FSPE+PS outperforms
  PS in a large class of heavy-tailed workloads} where SRPTE and FSPE
suffer. The net result is that a size-based policy such as FSPE+PS is
outperformed by PS only in extreme cases where \emph{both} the job size
distribution is extremely skewed \emph{and} job size estimation is
very imprecise.

It may appear surprising that, when job size skew is not extreme,
size-based scheduling can outperform PS even when size estimation
is very imprecise: even a small correlation between job size
and its estimation can direct the scheduler towards choices that are
beneficial on aggregate. In fact, as we see more in detail in the
following, sub-optimal scheduling choices become less penalized as the
job size skew diminishes.

\begin{figure}[!t]
  \centering
  \includegraphics[width=\plotwidth]{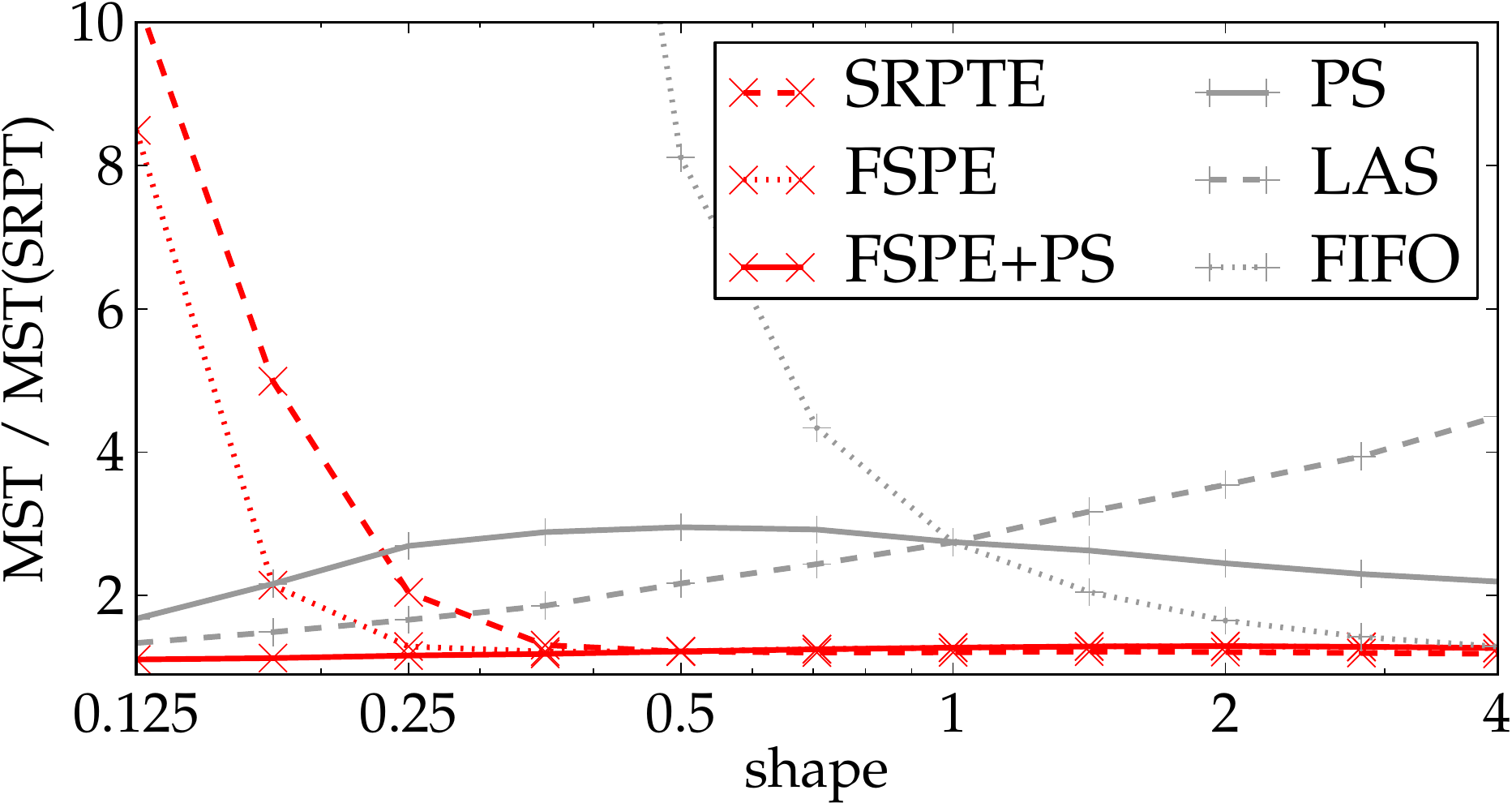}
  \caption{Impact of shape.}
  \label{fig:shape}
\end{figure}

\vspace{2mm}
\noindent
{\bf Impact of shape:} 
We now delve into details and examine how schedulers perform when
compared to the optimal MST that SRPT obtains. In the following Figures,
we show the ratio between the MST obtained with
the scheduling policies we implemented and the optimal one of SRPT.

From Figure~\vref{fig:shape}, we see that the shape parameter is
fundamental for evaluating scheduler performance. We notice that
FSPE+PS has \emph{almost optimal performance for all shape values
  considered} with the default sigma=0.5, which corresponds to 
  a correlation coefficient between job size and it estimate of 0.9, 
  while SRPTE and FSPE perform
poorly for highly skewed workloads. Regarding non size-based policies,
PS is outperformed by LAS for heavy-tailed workloads (shape $< 1$) and
by FIFO for light-tailed ones having shape $> 1$ ; PS provides a reasonable trade-off when the
job size distribution is unknown.  When the job size distribution is
exponential (shape $= 1$), non size-based scheduling policies perform
analogously; this is a result which has been proven analytically (see
\eg the work by Harchol-Balter~\cite{harchol2009queueing} and the
references therein).    It is interesting to consider
  the case of FIFO: in it, jobs are scheduled in series, and the
  priority between jobs is not correlated with job size: indeed, the
  MST of FIFO is equivalent to the one of a random scheduler executing
  jobs in series~\cite{klugman2012loss}. FIFO can be therefore seen as
  the limit case for a size-based scheduler such as FSPE or SRPTE when
  estimations carry no information at all about job sizes; the fact
  that errors become less critical as skew diminishes can be therefore
  explained with the similar patterns observed for FIFO.

\begin{figure}[!t]
  \subfloat[shape=0.25]{
    \includegraphics[width=\plotwidth]{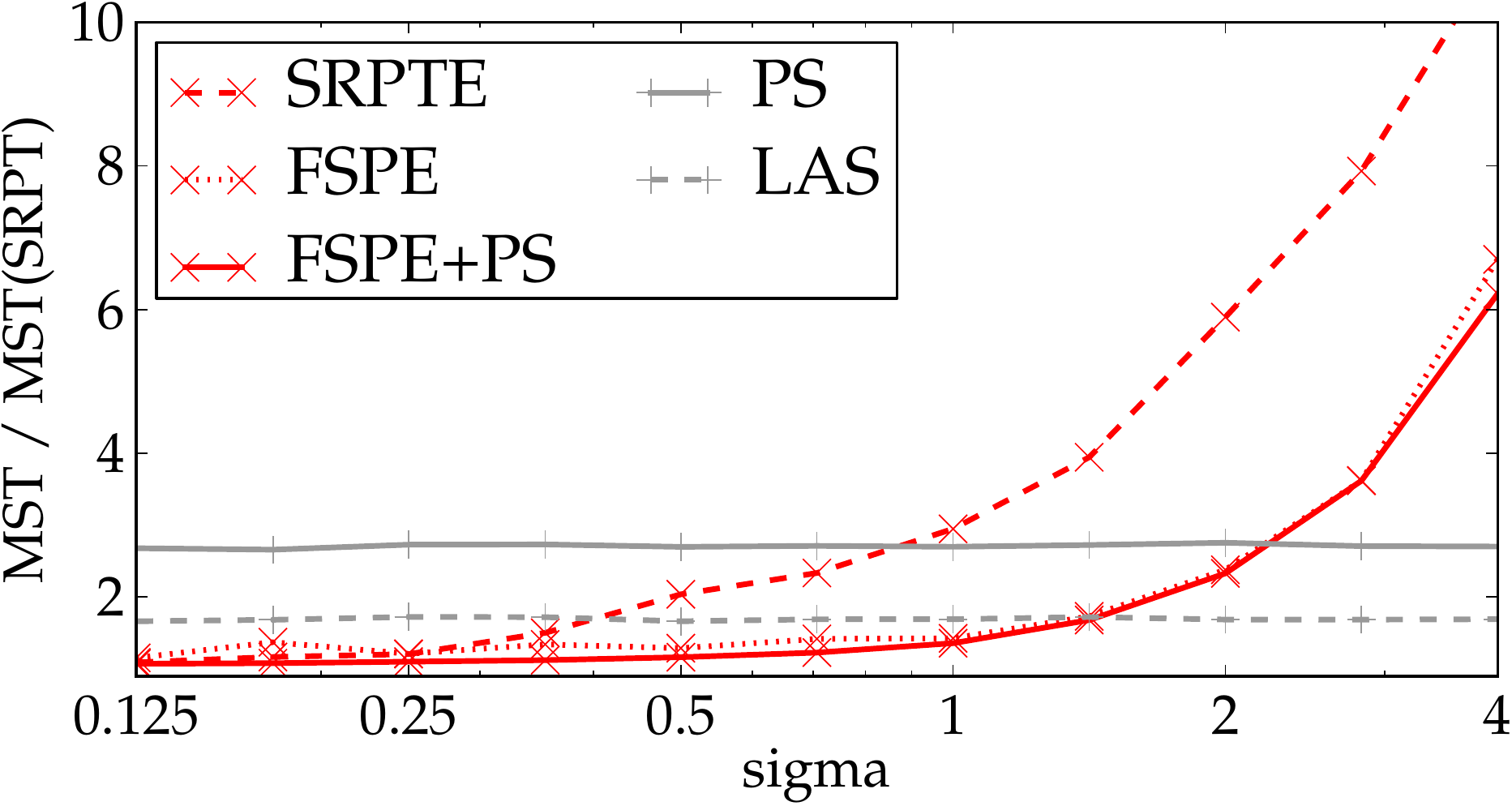}
    \label{fig:shape_0250}
  }
  \\
  \subfloat[shape=0.177]{
    \includegraphics[width=\plotwidth]{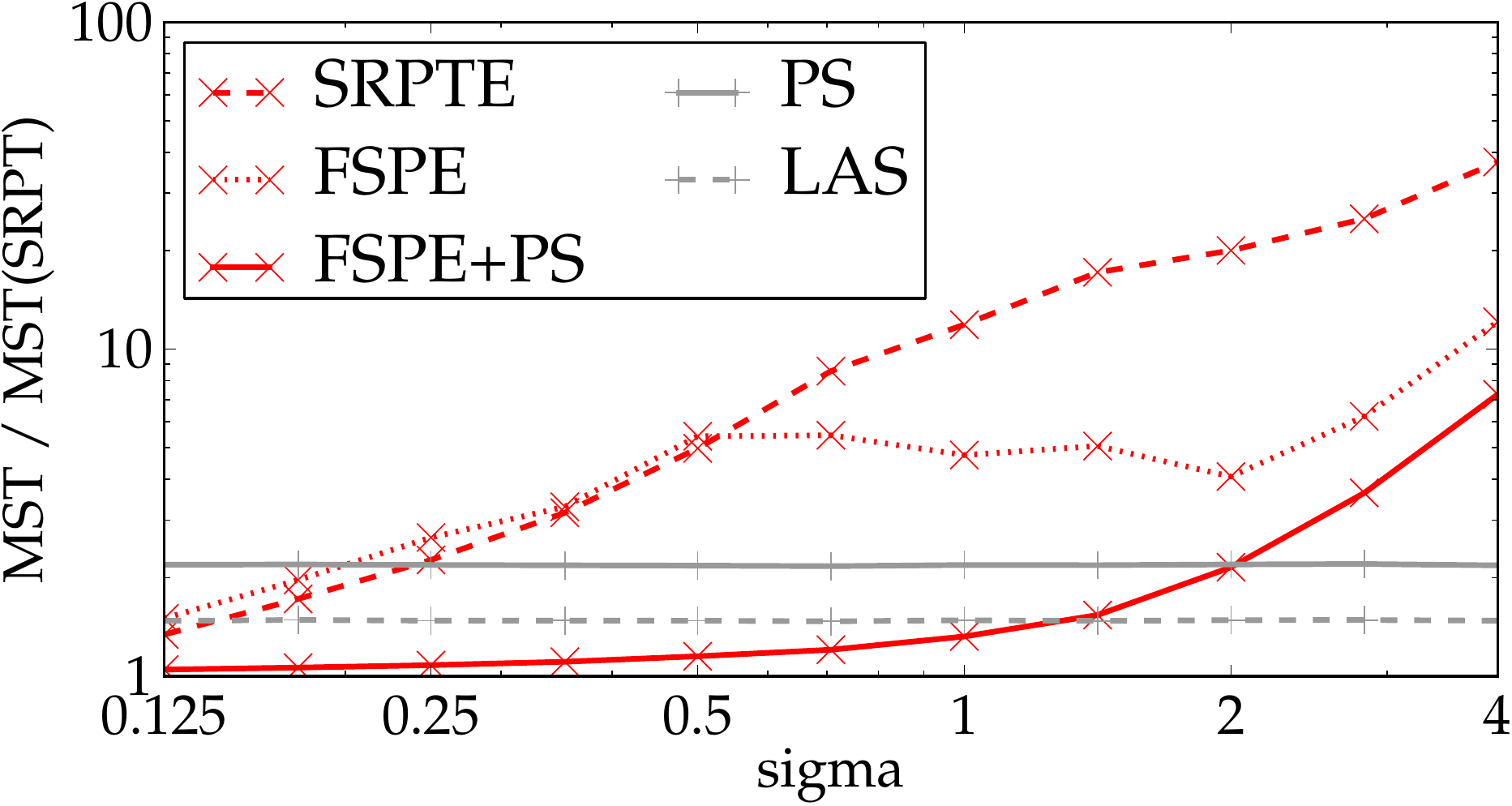}
    \label{fig:shape_0177}
  }
  \\
  \subfloat[shape=0.125]{
    \includegraphics[width=\plotwidth]{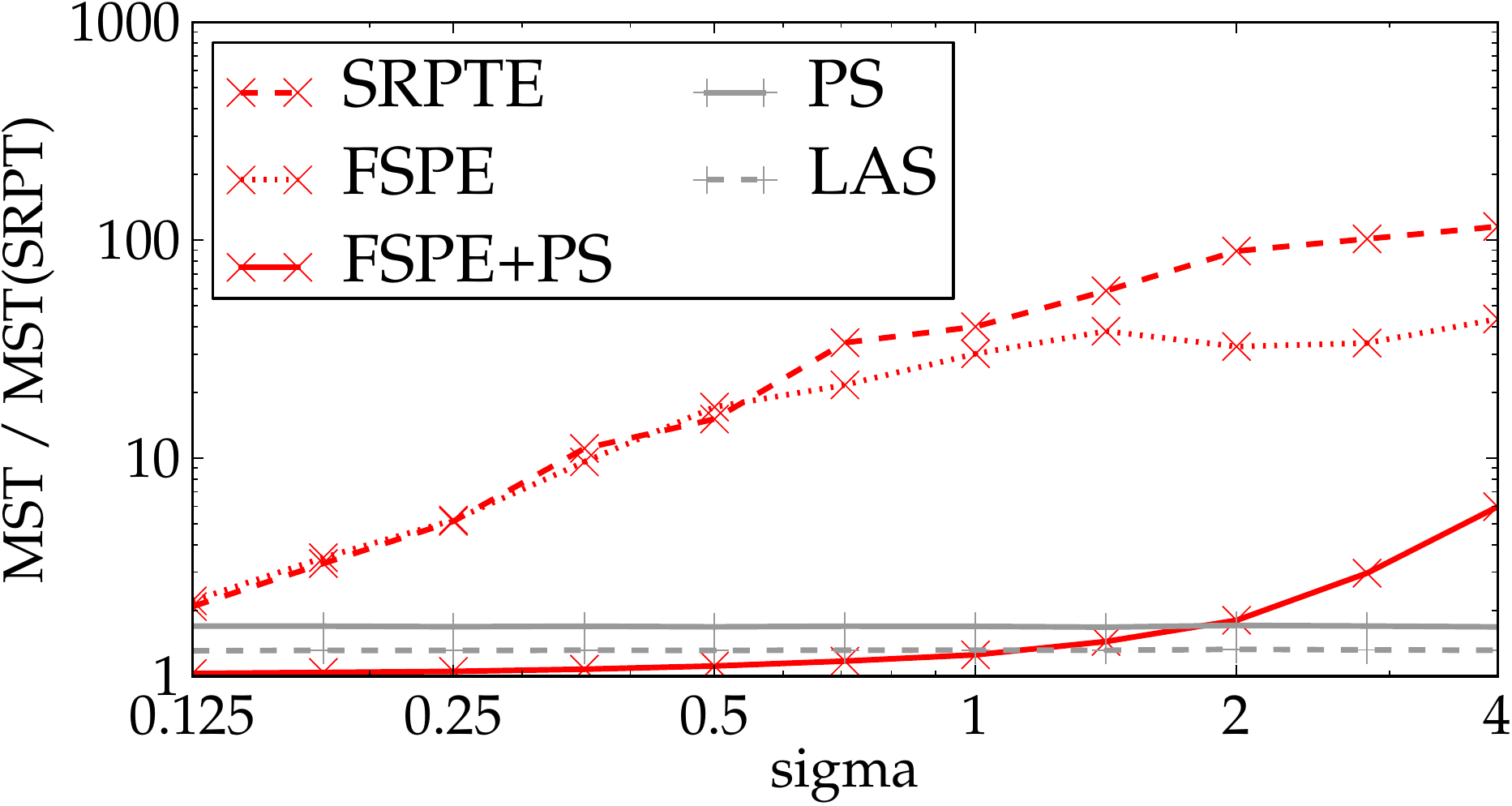}
    \label{fig:shape_0125}
  }
  \caption{Impact of error on heavy-tailed workloads, sorted by growing skew.}
  \label{fig:mst_heavytail}
  \label{fig:sigma}
\end{figure}

\vspace{2mm}
\noindent
{\bf Impact of sigma:}
The shape of the job size distribution is
fundamental in determining the behavior of scheduling algorithms, and
heavy-tailed job size distributions are those in which the
behavior of size-based scheduling differs noticeably. Because of this,
and since heavy-tailed workloads are central in the literature on
scheduling, we focus on those.

In Figure~\vref{fig:sigma}, we show the impact of the sigma parameter
representing error for three heavily skewed workloads. In all three
plots, the values for FIFO fall outside of the plot. These plots
demonstrate that FSPE+PS is robust with respect to errors in all the
three cases we consider, while SRPTE and FSPE suffer as the skew
between job sizes grows. In all three cases, FSPE+PS performs better
than PS as long as sigma is lower than 2: this
corresponds to lax bounds on size estimation quality,
  requiring a correlation coefficient between job size and its
estimate of 0.15 or more.

In all three plots, FSPE+PS performs better than SRPTE; the difference
between FSPE+PS and FSPE, instead, becomes discernible only for
$\textrm{shape}<0.25$. We explain this difference by noting that, when
several jobs are in the queue, size reduction in the virtual queue of
FSPE is slow: this leads to less jobs being late and therefore non
preemptable. As the distribution becomes more heavy-tailed, more jobs
become late in FSPE and differences between FSPE and FSPE+PS become
significant, reaching differences of even around one order of
magnitude.

In particular in Figure~\ref{fig:shape_0177}, there are areas ($0.5 <
\mathrm{sigma} < 2$) in which increasing errors decreases (slightly)
the MST of FSPE.  This counterintuitive phenomenon is explained by the
characteristics of the error distribution: the mean of the log-normal
distribution grows as sigma grows, therefore the aggregate amount of
work for a set of several jobs is more likely to be over-estimated;
this reduces the likelihood that several jobs at once become late and
therefore non-preemptable. In other words, FSPE works better with
estimation means that tend to over-estimate job size; however, it is
always better to use FSPE+PS, which provides a more reliable and
performant solution to the same problem.

\begin{figure}[!t]
  \subfloat[$\alpha=2$]{
    \includegraphics[width=\plotwidth]{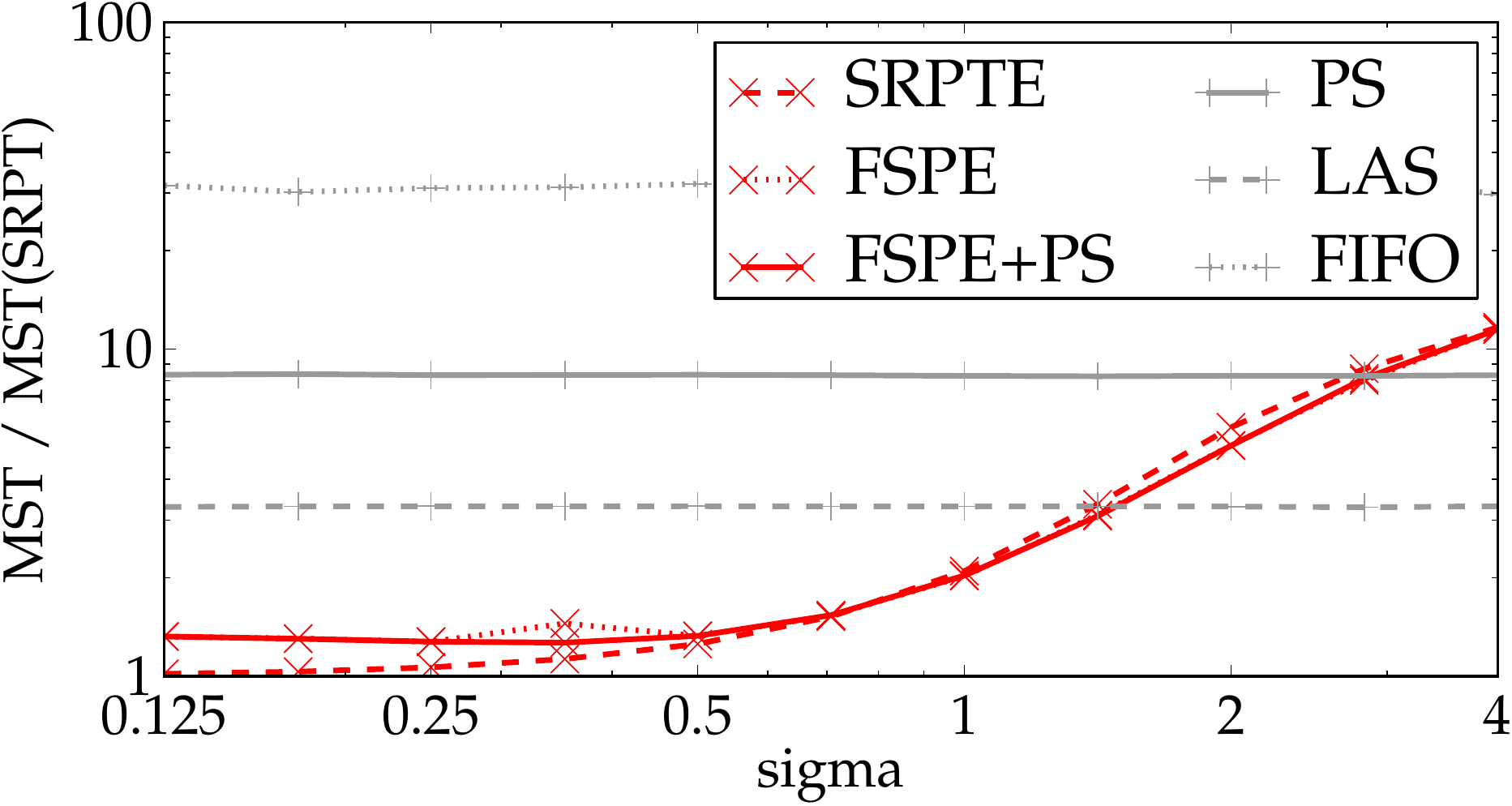}
    \label{fig:pareto2}
  }
  \\
  \subfloat[$\alpha=1$]{
    \includegraphics[width=\plotwidth]{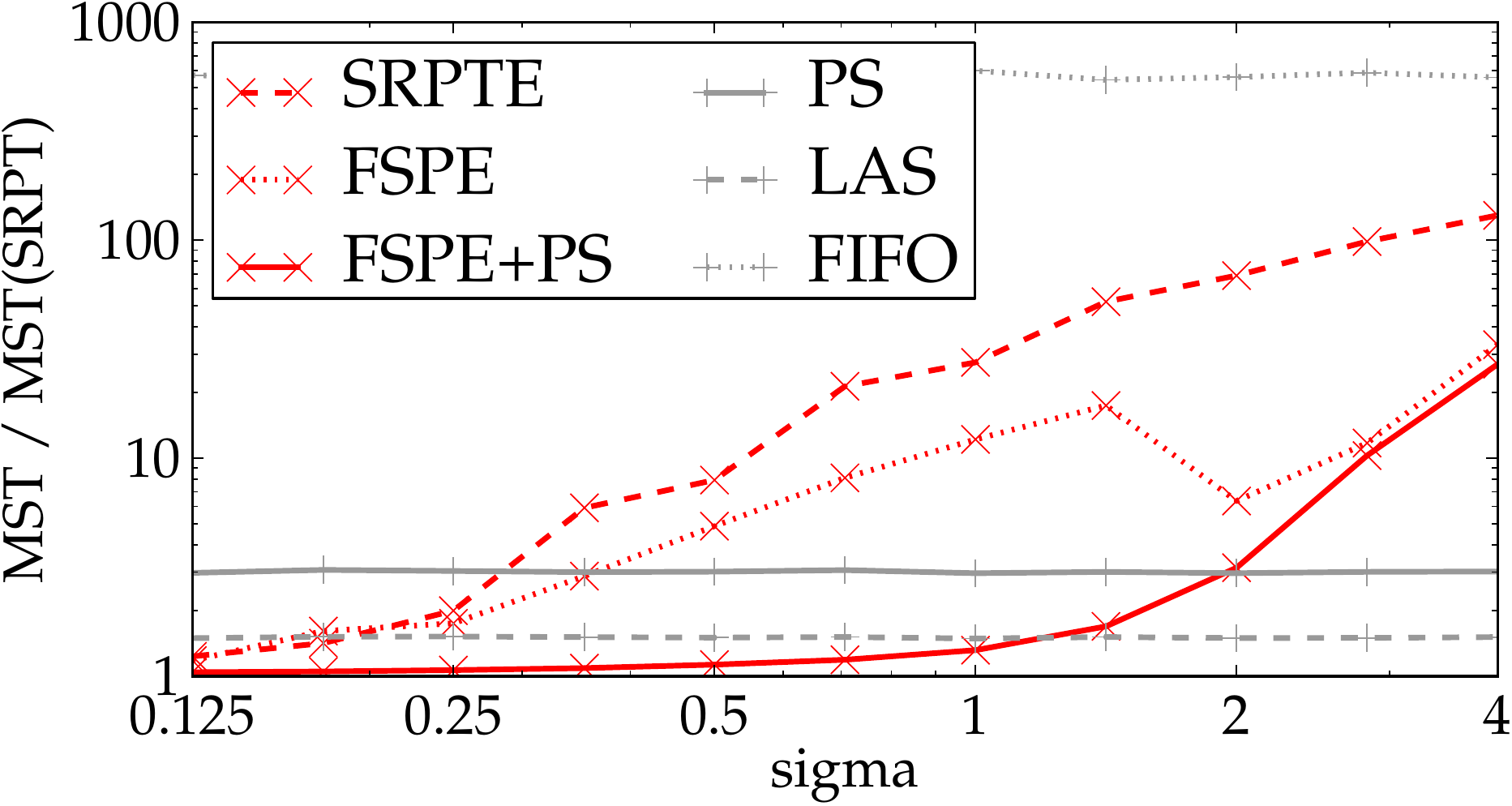}
    \label{fig:pareto1}
  }
  \caption{Pareto job size distributions, sorted by growing skew.}
  \label{fig:pareto}
\end{figure}

\vspace{2mm}
\noindent
{\bf Pareto Job Size Distribution:} In the literature, workloads are
often generated using the Pareto distribution. To help comparing our
results to the literature, in Figure~\vref{fig:pareto} we show results
for job sizes having a Pareto distribution, using $x_m=0$ and
$\alpha=\{1,2\}$. The results we observe for the Weibull distribution
are still qualitatively valid for the Pareto distribution; the value
of $\alpha=1$ is roughly comparable to a shape of 0.15 for the Weibull
distribution, while $\alpha=2$ is comparable to a shape of around 0.5,
where the three size-based disciplines we take into account still have
similar performance.

\begin{figure}[!t]
  \subfloat[Varying load.]{
    \includegraphics[width=\plotwidth]{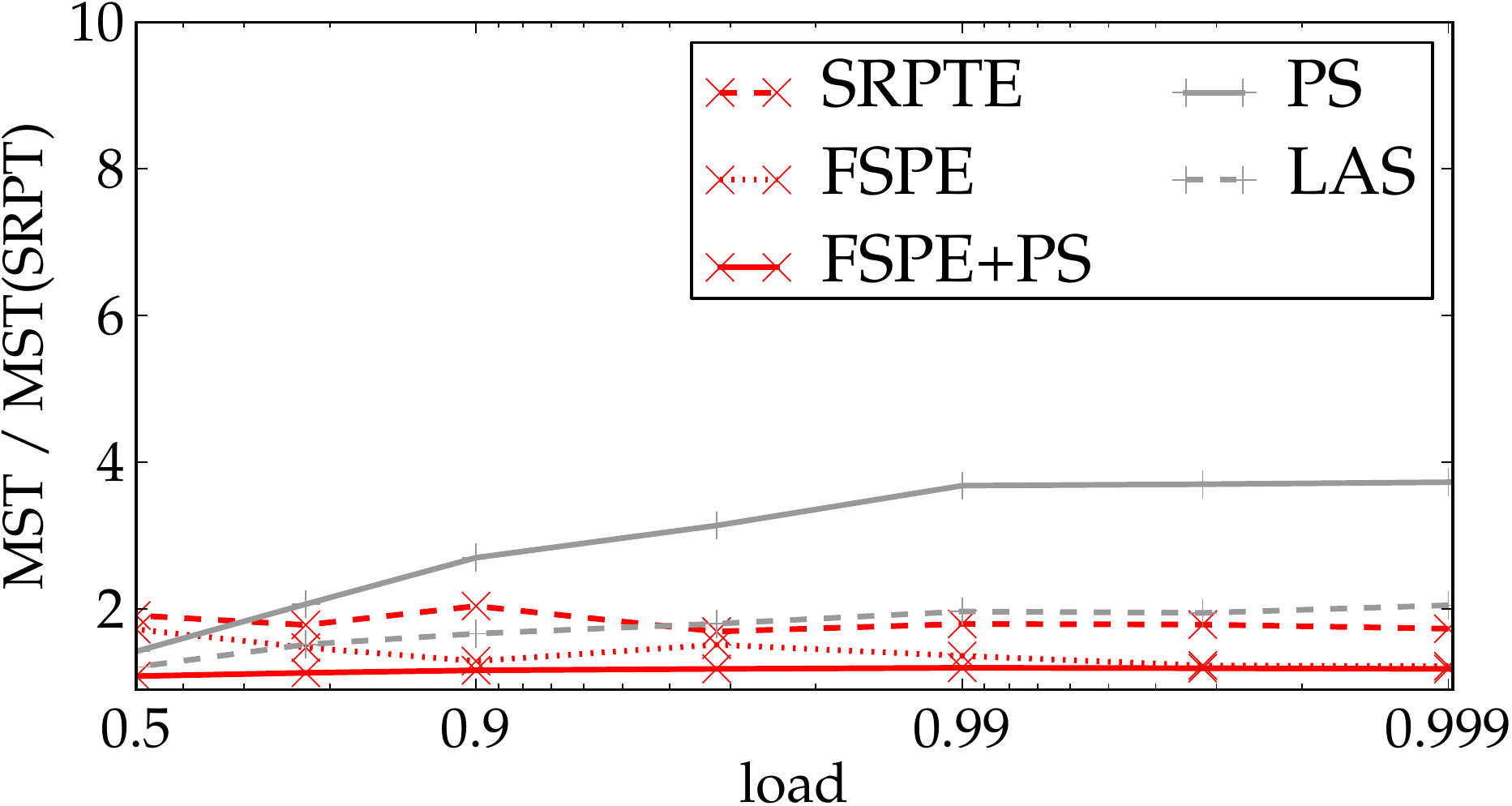}
    \label{fig:load}
  }
  \\
  \subfloat[Varying timeshape.]{
    \includegraphics[width=\plotwidth]{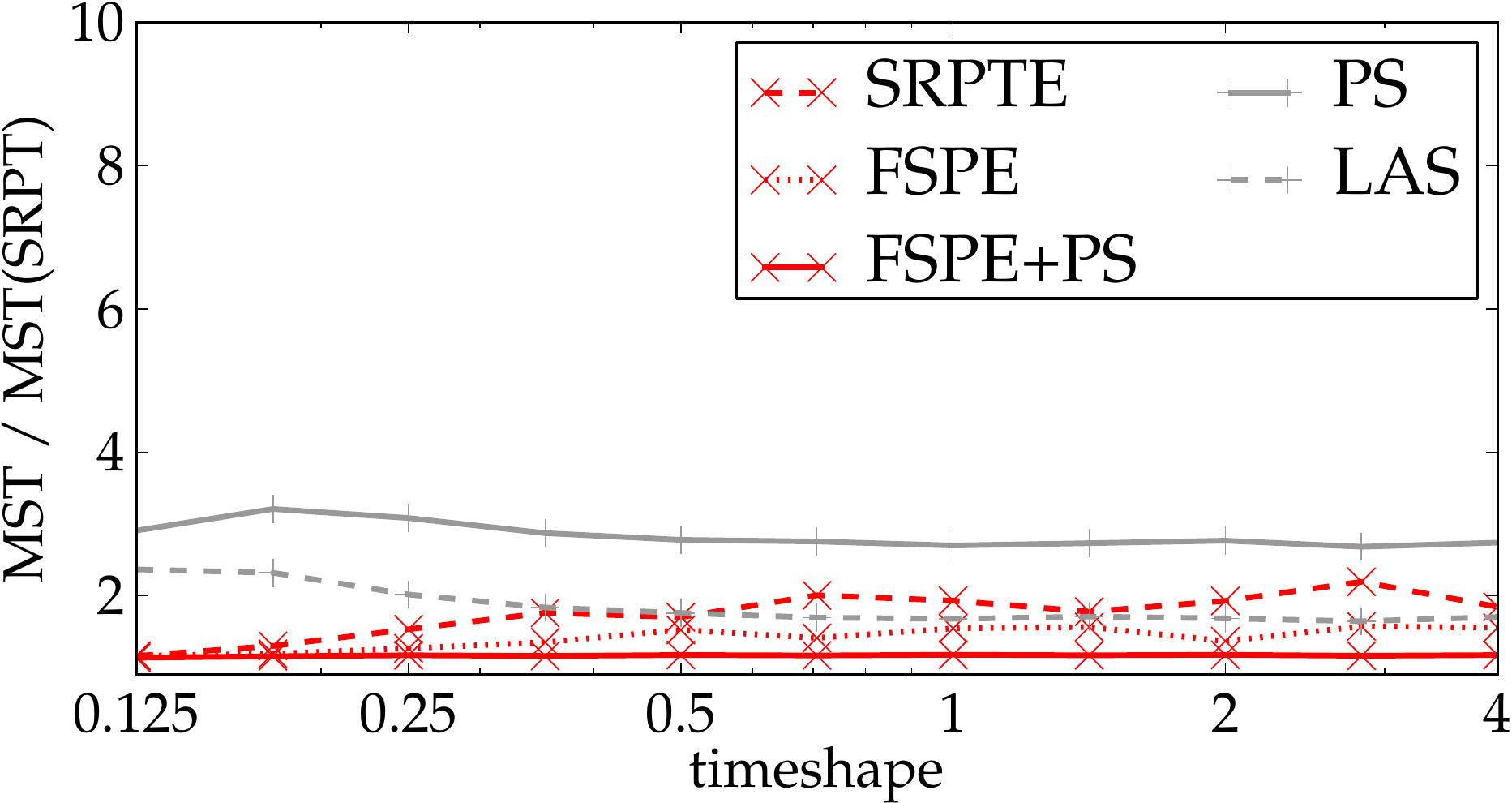}
    \label{fig:timeshape}
  }
  \caption{Impact of load and timeshape.}
  \label{fig:sim}
\end{figure}

\vspace{2mm}
\noindent
{\bf Impact of Other Parameters:}
In Figure~\vref{fig:sim}, we show the impact of varying load and
timeshape, while keeping sigma and shape at their default
values.

Figure~\ref{fig:load} shows that performance of size-based scheduling
protocols is not heavily impacted by load, as the ratio between the
MST obtained and the optimal one remains roughly constant (note that
the graph shows a ratio, not the absolute values, that increase as the 
load increases); conversely,
non size-based schedulers such as PS and LAS deviate more from optimal
as the load grows.

Figure~\ref{fig:timeshape} shows the impact of changing the timeshape
parameter: with low values of timeshape, job submissions are bursty
and separated by long pauses; with high values job submissions are
evenly spaced. We note that size-based scheduling policies respond
very well to bursty submissions where several jobs are submitted at
once: in this case, adopting a size-based policy that focuses all the
system resources on the smallest jobs pays best; as the intervals
between jobs become more regular, SRPTE and FSPE become slightly less
performant; FSPE+PS remains close to optimal.




\begin{figure}[!t]
  \centering
  \includegraphics[width=\plotwidth]{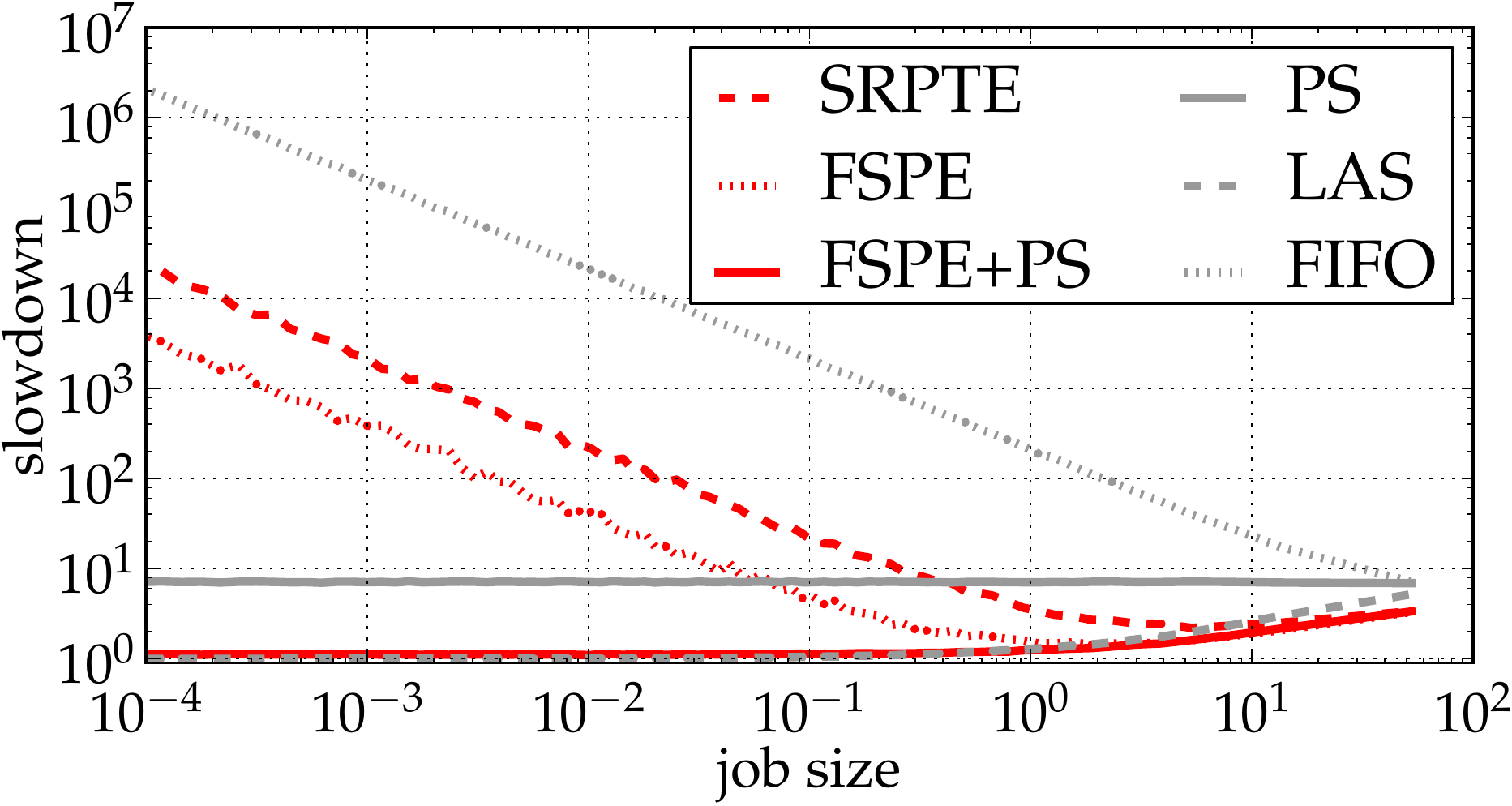}
  \caption{Mean conditional slowdown.}
  \label{fig:size_vs_slowdown}
\end{figure}

\vspace{2mm}
\noindent
{\bf Conditional Slowdown:}
We now consider the topic of fairness, intending here -- as discussed
in Section~\ref{sec:metrics} -- that jobs' running time should be
proportional to their size, and therefore not experience large
slowdowns.

To better understand the reason for the unfairness of FIFO, SRPTE and
FSPE, in Figure~\vref{fig:size_vs_slowdown} we evaluate \emph{mean
 conditional slowdown}, comparing job size with the average slowdown
(job sojourn time divided by job size) obtained at that size using our
default simulation parameters. The figure has been obtained by sorting
jobs by size and binning them in 100 equally sized classes of jobs
with similar size; points plotted are obtained by averaging job size
and slowdown in each of the 100 class.

The almost parallel lines of FIFO, SRPTE and FSPE for smaller jobs are
explained by the fact that, below a certain size, \emph{job sojourn
  time is essentially independent from job size}: indeed, it depends
on the total size of older (for FIFO) or late (for SRPTE and FSPE)
jobs at submission time.

We confirm experimentally the fact that the expected slowdown in PS is
constant, irrespectively of job size~\cite{wierman2007fairness};
FSPE+PS and LAS, on the other hand, have close to optimal slowdown for
small jobs. The better MST of FSPE+PS is instead due to better
performance for larger jobs, which are more penalized in LAS.

\begin{figure}[!t]
  \centering
  \subfloat{\includegraphics[width=\plotwidth]{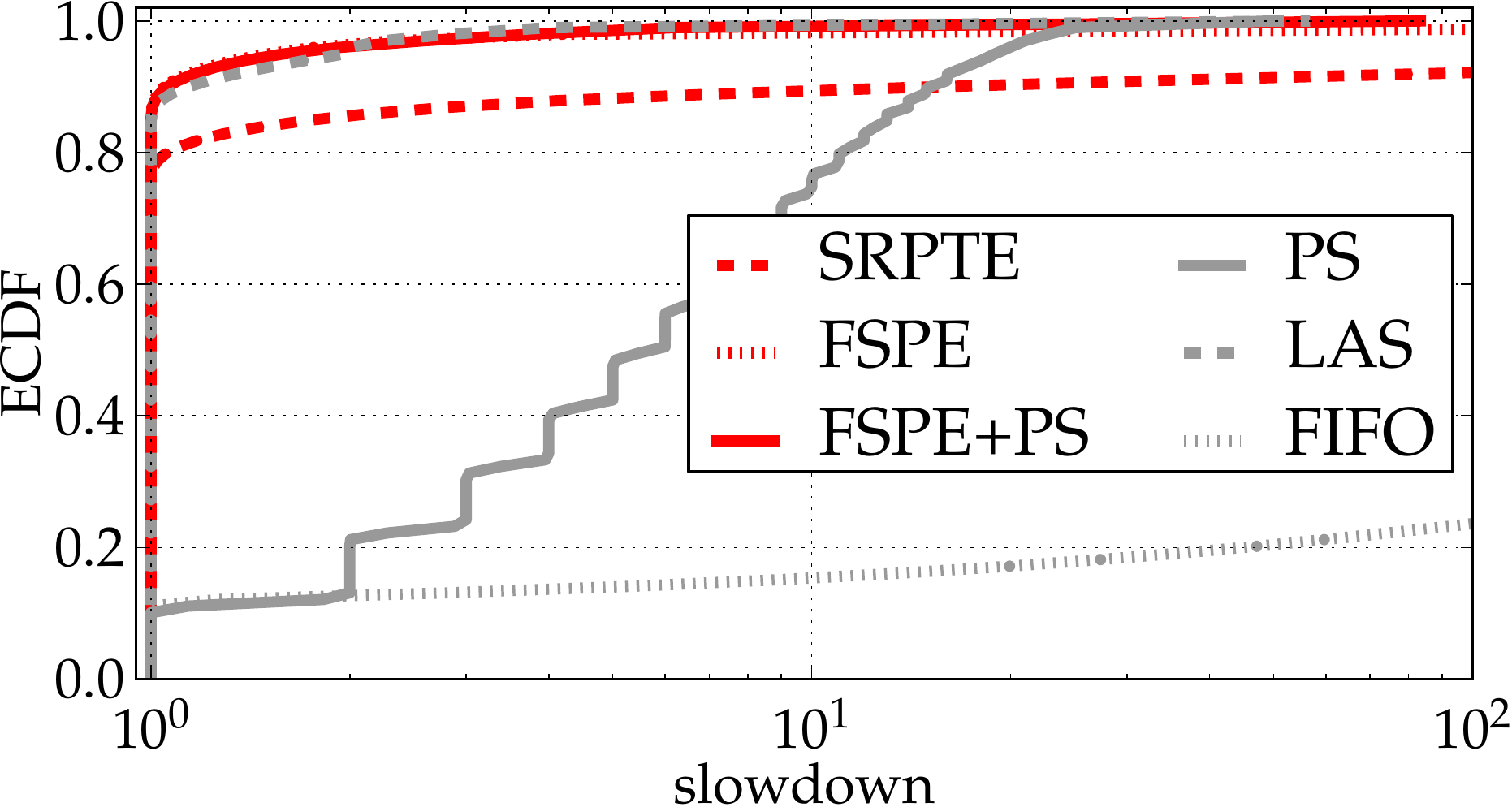}}
  \\
  \subfloat{\includegraphics[width=\plotwidth]{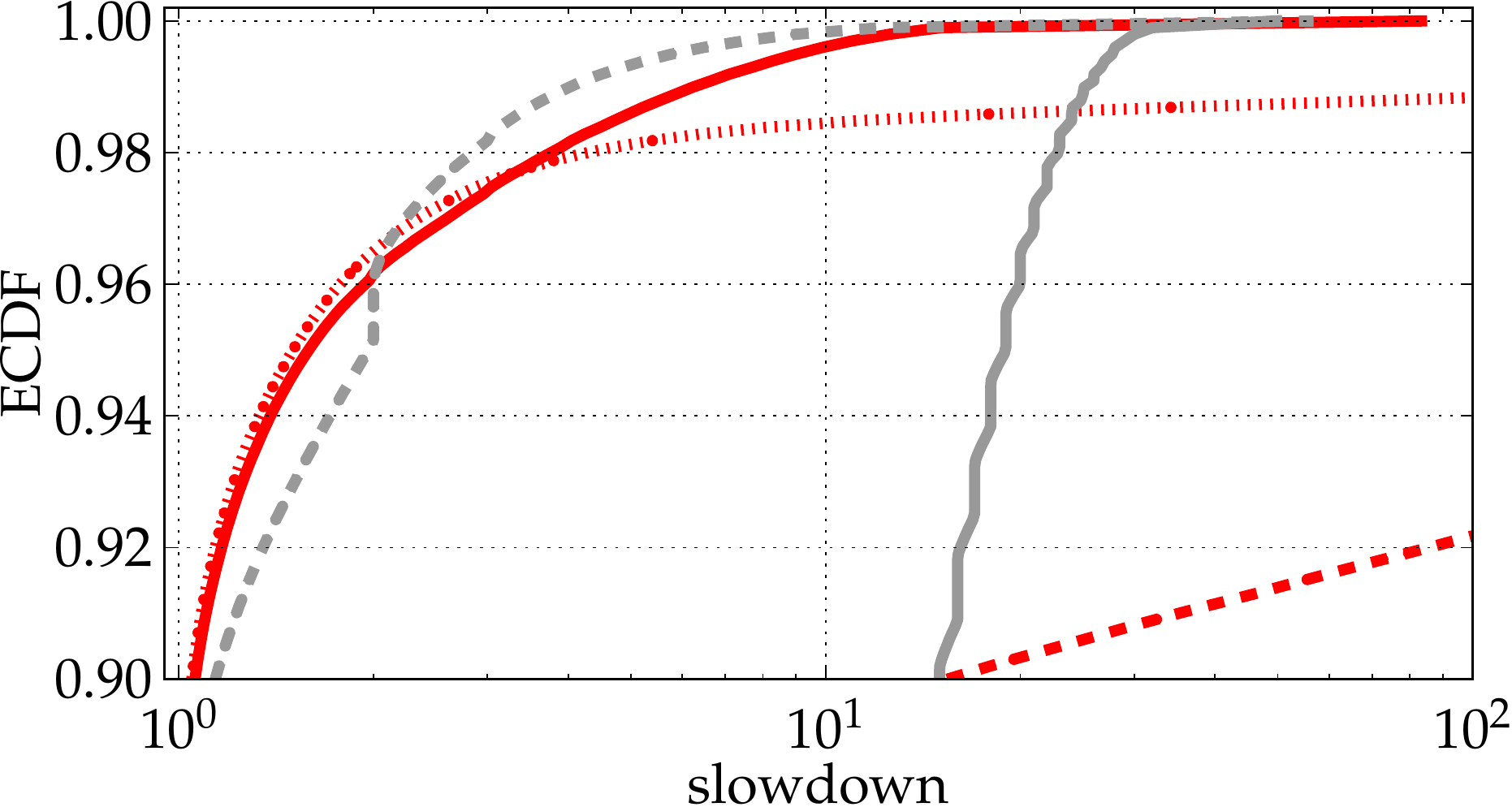}}
  \caption{Per-job slowdown: full CDF (top) and zoom on the 10\% more
    critical cases (bottom).}
  \label{fig:slowdown}
\end{figure}

\vspace{2mm}
\noindent
{\bf Per-Job Slowdown:}
The results we have shown testify that, for FSPE+PS and similarly to
LAS, slowdown values are homogeneous across classes of job sizes:
neither small nor big jobs are penalized when using FSPE+PS. This is a
desirable result, but the reported results are still averages: in
order to ensure that sojourn time is commensurate to size \emph{for
  all jobs}, we need to investigate the \emph{per-job} slowdown distribution.

In Figure~\vref{fig:slowdown}, we plot the CDF of per-job slowdown for
our default simulator parameters.  By serving efficiently smaller
jobs, all size-based scheduling techniques and LAS manage to obtain an
optimal slowdown of 1 for the majority of jobs. However, some jobs
experience very high slowdown values: jobs with a slowdown larger than
100 are around 1\% for FSPE and around 8\% for SRPTE.

PS, LAS, and FSPE+PS perform well in terms of fairness, with no jobs
experiencing slowdown higher than 100 in our experiment
runs.\footnote{Figure~\ref{fig:slowdown} plots the results of 121
  experiment runs, representing therefore 1,210,000 jobs in this
  simulation.} While PS is generally considered the reference for a
``fair'' scheduler, it obtains slightly better slowdown than LAS and
FSPE+PS only for the most extreme cases, while being outperformed for
a large majority of the jobs. We remark that slowdown values for PS
are clustered around integer values, because they are obtained in the
common case where a small job is submitted when $n$ larger ones are
running.

\subsection{Real Workloads}
\label{sec:real_workloads}

We now consider two real workloads in order to confirm that the
phenomena we observed in our experiments are not an artifact of the
synthetic traces that we generated, and that they indeed apply in
realistic cases. From the traces we obtain two data points per job:
submission time and job size. In this way, we move away from the
assumptions of the $G/G/1$ model, and we provide results that can
account for more general cases where periodic patterns and correlation
between job size and submission times are present.

\begin{figure}[!t]
  \centering
  \includegraphics[width=\plotwidth]{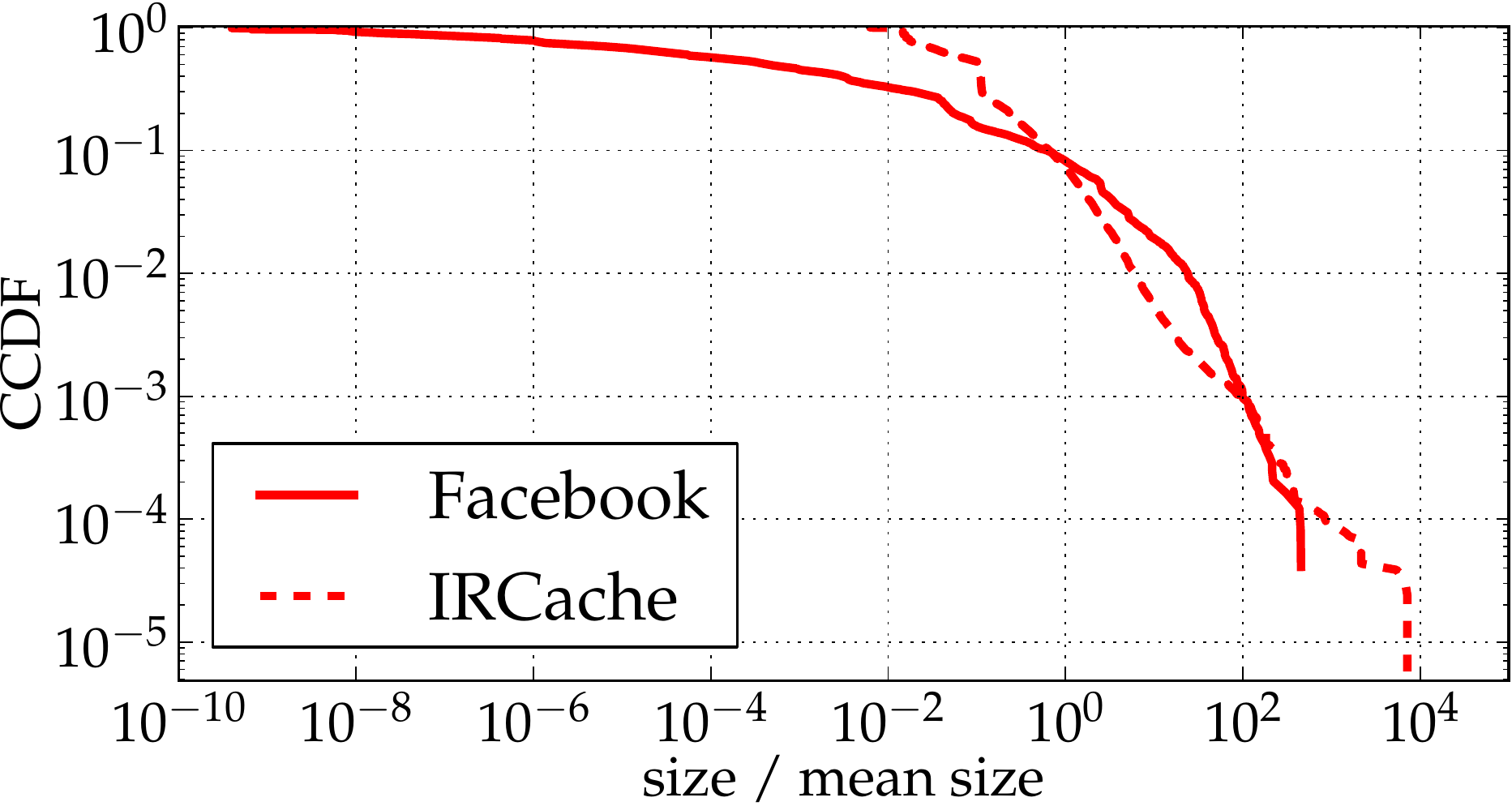}
  \caption{CCDF for the real workloads.}
  \label{fig:ccdf}
\end{figure}

\vspace{2mm}
\noindent
{\bf Hadoop at Facebook:} We consider a trace from a Facebook Hadoop
cluster in 2010, covering one day of job submissions. The trace has
been collected and analized by Chen \etal~\cite{chen2012interactive};
it is comprised of 24,443 jobs and it is available
online.\footnote{\url{https://github.com/SWIMProjectUCB/SWIM/blob/master/workloadSuite/FB-2010_samples_24_times_1hr_0.tsv}}
For the purposes of this work, we consider the job size as the number
of bytes handled by each job (summing input, intermediate output and
final output): the mean size is 76.1 GiB, and the largest job
processes 85.2 TiB. To understand the shape of the tail for the job
size distribution, in Figure~\vref{fig:ccdf} we plot the complementary
CDF (CCDF) of job sizes (normalized against the mean); the
distribution is heavy-tailed and the largest jobs are around 3 orders
of magnitude larger than the average size. For homogeneity with the
results of Section~\ref{sec:synthetic}, we set the processing speed of
the simulated system (in bytes per second) in order to obtain a load
(total size of the submitted jobs divided by total length of the
submission schedule) of 0.9.

\begin{figure}[!t]
  \centering
  \includegraphics[width=\plotwidth]{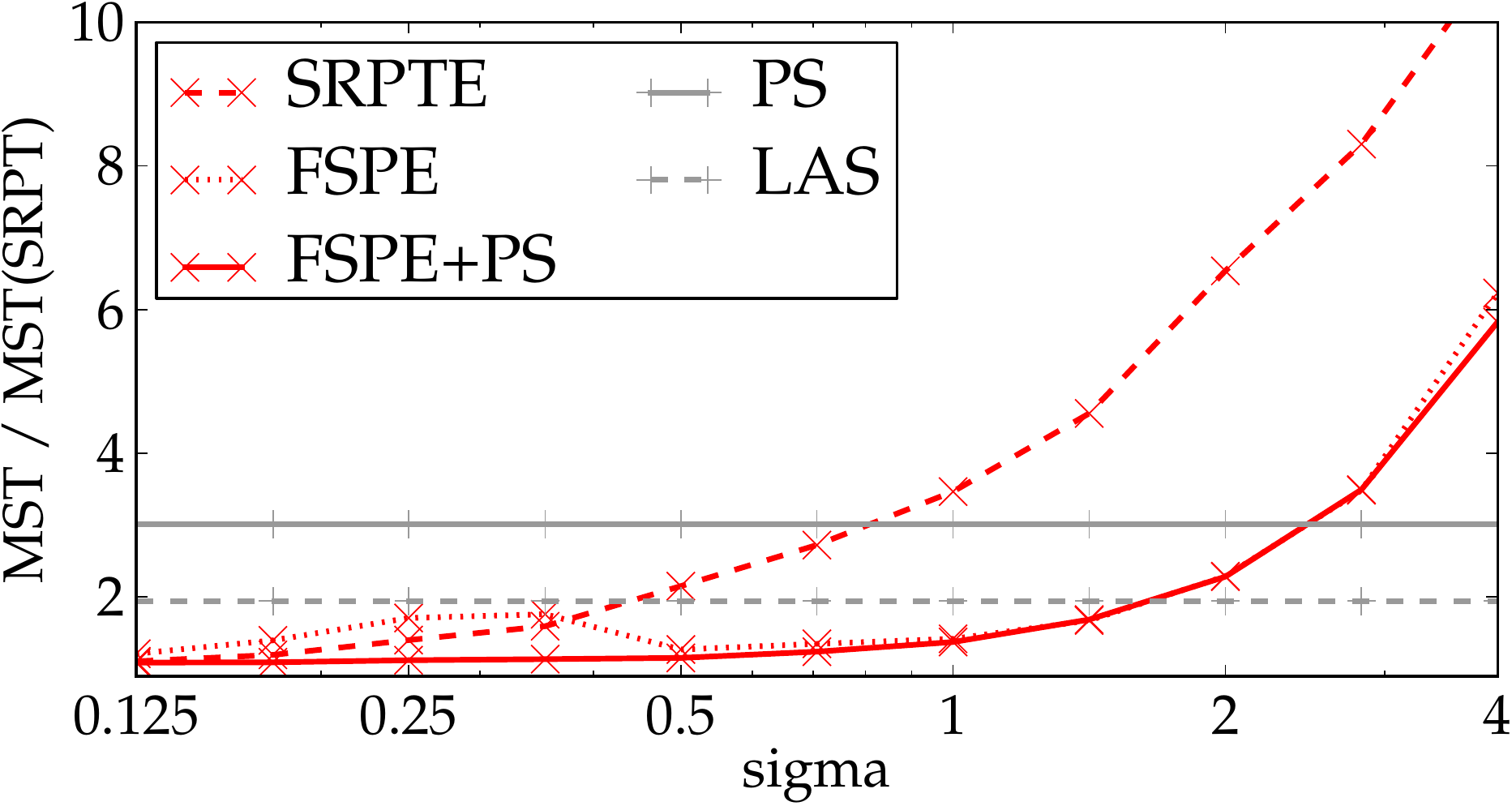}%
  \caption{MST of the Facebook workload.}
  \label{fig:fb}
\end{figure}

In Figure~\vref{fig:fb}, we show MST, normalized against optimal MST,
while varying the error rate. We remark that these results are very
similar to those that we observe from Figure~\vref{fig:sigma}: also in
this case, FSPE and FSPE+PS perform well even when job size estimation
errors are far from negligible. We do not plot the
  slowdown CDF for space limitations: results are analogous to those
  observed in Figure~\vref{fig:slowdown}.  These results show that
this workload is well represented by our synthetic workloads, when
shape is around 0.25.

We performed more experiments on these traces; extensive results are
available in a technical report~\cite{dell2013simulator}.

\vspace{2mm}
\noindent
{\bf Web Cache:} IRCache (\href{http://ircache.net}{ircache.net}) is a research
project for web caching; traces from the caches are freely
available. We performed our experiments on a one-day trace of a server
from 2007 totaling 206,914
requests;\footnote{\url{ftp://ftp.ircache.net/Traces/DITL-2007-01-09/pa.sanitized-access.20070109.gz}.}
the mean request size in the traces is 14.6KiB, while the maximum
request size is 174 MiB. In Figure \vref{fig:ccdf} we show the CCDF of
job size; as compared to the Facebook trace analyzed previously, the
workload is more heavily tailed: the biggest requests are four orders
of magnitude larger than the mean. As before, we set the simulated
system processing speed in bytes per second to obtain a load of 0.9.

\begin{figure}[!t]
  \centering
  \includegraphics[width=\plotwidth]{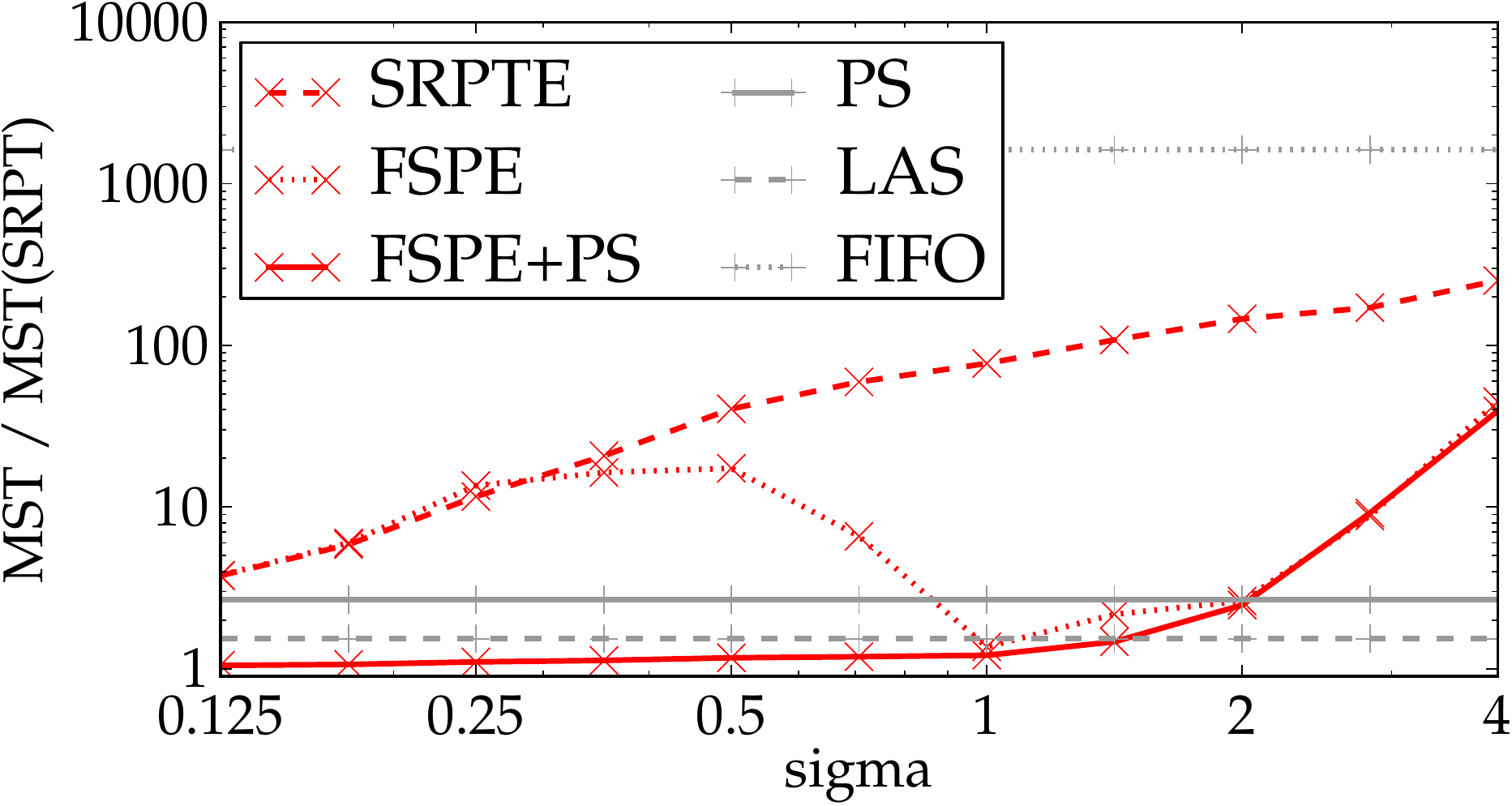}
  \caption{MST of the IRCache workload.}
  \label{fig:webserver}
\end{figure}

In Figure~\vref{fig:webserver} we plot the evolution of MST as the
sigma parameter controlling error grows. Since the job size
distribution is more heavily tailed, sojourn times are more influenced
by job size estimation errors (notice the logarithmic scale on the $y$
axis), confirming the results we have from Figure~\vref{fig:3d}. The
performance of FSPE does not worsen monotonically as error grows, but
rather becomes better for $0.5 < \mathrm{sigma} < 1$; this is a
phenomenon that we also observe -- albeit to a lesser extent -- for
synthetic workloads in Figure~\vref{fig:shape_0177} and for the
Facebook workload in Figure~\vref{fig:fb}. The explanation that we
provided in Section~\ref{sec:synthetic} applies: since the mean of the
log-normal distribution grows as sigma grows, the aggregate amount of
work for a given set of jobs is likely to be over-estimated in total,
reducing the likelihood that several jobs at once become late and
therefore non-preemptable. Also in this case, we still remark that
FSPE+PS consistently outperforms FSPE. Once again, the
  results for the slowdown distribution -- not reported for space
  limitations -- are qualitatively analogous to those reported in
  Section~\ref{sec:exp_slowdown}.

\balance

\section{Conclusion}
\label{sec:conclusion}

This work shows that size-based scheduling is an applicable and
performant solution in a wide variety of situations where job size is
known approximately rather than exactly. The limitations shown by
previous work are, in a large part, solved by the approach we took in
FSPE+PS, which is a simple modification to FSP; analogous measures can
be taken in other preemtpive size-based scheduling disciplines.

FSPE+PS also solves a fairness problem: while FSPE and SRPTE penalize
small jobs and results in slowdown values which are not proportionate
to their size, FSPE+PS has constant and optimal slowdown for most
small jobs.
Our work suggests that, if even rough estimates can be produced to
estimate job sizes, it is worthy to try size-based scheduling: our
proposal, FSPE+PS, is essentially as complex to implement as FSP is,
and provides close to optimal response times and good fairness in all
but the most extreme of cases.

We released our simulator as free software: it can be reused for
\textit{1)} reproducing our experimental results; \textit{2)}
prototype new scheduling algorithms; \textit{3)} predict system
behavior in particular cases, by replaying traces.

We are currently evaluating other scheduling variants, going beyond
FSPE+PS with the goal of having a scheduler that adapts dynamically to
the characteristics of job size and error distribution, striving to
\emph{always} be at least as performant as non size-based scheduling
policies such as LAS.

\section*{Acknowledgments}

This work has been partially supported by the EU projects BigFoot
(FP7-ICT-223850) and mPlane (FP7-ICT-318627).

\bibliographystyle{IEEEtran}
\bibliography{references}

\end{document}